\documentclass[11pt, reqno]{amsart}

\usepackage{amsfonts}
\usepackage{appendix}
\usepackage{amsmath, amsthm, amssymb}

\usepackage{ulem} 
\usepackage{color}
\usepackage[dvips,left=1in, right=1in, top=1in, bottom=1in]{geometry}

\newcounter{notenum}

\newcommand{\argmax}{\operatorname{argmax}}

\newcommand{\esssup}{\operatorname{esssup}}
\newcommand{\essinf}{\operatorname{essinf}}

\providecommand{\R}{} \renewcommand{\R}{{\mathbb R}}

\newcommand{\N}{{\mathbb N}}
\newcommand{\PP}{{\mathbb P}}
\newcommand{\QQ}{{\mathbb Q}}

\newcommand{\EE}{{\mathbb E}}
\newcommand{\FF}{{\mathcal F}}

\newcommand{\MM}{{\mathcal M}}
\newcommand{\NN}{{\mathcal N}}

\newcommand{\EN}{{\mathcal E}}

\renewcommand{\AA}{{\mathcal A}}
\newcommand{\PPP}{{\mathcal P}}

\newcommand{\el}{{\mathbb L}} 

\newcommand{\lone}{\el^1}

\newcommand{\linf}{\el^{\infty}}

\newcommand{\phione}{\phi}
\newcommand{\phitwo}{\psi}
\newcommand{\tone}{\theta}
\newcommand{\ttwo}{\vartheta}

\newcommand{\gone}{\gamma}
\newcommand{\gtwo}{g}
\newcommand{\Ai}{A^{i}}

\newtheorem{theorem}{Theorem}[section]
\theoremstyle{plain}

\newtheorem{assumption}{Assumption}[section]

\newtheorem{corollary}{Corollary}[section]
\newtheorem{definition}{Definition}[section]
\newtheorem{example}{Example}[section]
\newtheorem{lemma}{Lemma}[section]
\newtheorem{proposition}{Proposition}[section]
\newtheorem{remark}{Remark}[section]

\begin{document}
\title{Forward Exponential Performances:
Pricing and Optimal Risk Sharing} \maketitle

\ \\[0.3ex]

\begin{center}
\begin{minipage}{0.8\textwidth}
\begin{center}
{\bf\large Michail Anthropelos}\\
Department of
Banking and Financial Management\\
University of Piraeus\\
{\tt anthropel@unipi.gr}\\
\end{center}
\end{minipage}
\end{center}

\ \\[0.2ex]
\begin{quote}
\noindent{\bf Abstract.} In a Markovian stochastic volatility
model, we consider financial agents whose investment criteria are
modelled by forward exponential performance processes. The problem
of contingent claim indifference valuation is first addressed and
a number of properties are proved and discussed. Special attention
is given to the comparison between the forward exponential and the
backward exponential utility indifference valuation. In addition,
we construct the problem of optimal risk sharing in this forward
setting and solve it when the agents' forward performance criteria
are exponential.
\end{quote}

\ \\[0.2ex]\textbf{Key words.} Forward performance criteria,
stochastic utility, stochastic risk aversion, indifference price,
optimal risk sharing, exponential utility, contingent claim
pricing\\


\ \\[0.5ex]

\section*{Introduction}

Contingent claim pricing in incomplete markets is one of the most
challenging problems in mathematical finance. In incomplete
markets, there exist contingent claims for which there is no
dynamic self-financing portfolio that perfectly replicates their
payoffs. A consequence of this is that the non-arbitrage arguments
provide only an interval of prices consistent with the
non-arbitrage assumption. The answer to the question which is the
``\textit{correct}'' price within this interval requires a model
of agents' risk preferences and perhaps their endowments or/and
their beliefs.

One of the most fruitful literature on financial agents' risk
preferences is the one on utility function. Based on the work of
R.~Merton \cite{Mert69} and \cite{Mert71}, this theory suggests
that an agent, who models her risk preferences through a utility
function, is going to invest in the financial market with the aim
to maximize the expectation of her utility function. The
\textit{utility maximization} has been extensively studied and
developed in a variety of market models and utility functions (see
for instance, \cite{HePea91}, \cite{KarLehShrXu91},
\cite{KraSch03}  and \cite{Sch04} for an overview). If an agent's
investment criterion is the utility maximization, it is reasonable
to assume that she evaluates each contingent claim by comparing
the following two situations: maximization of the expected utility
after buying (selling) the claim and maximization of the utility
without any transaction on the claim. The price that makes these
situations indifferent for the agent's perspective is the
so-called \textit{indifference price}. This pricing mechanism was
introduced in mathematical finance literature in \cite{HodNeu89}
and then further developed by a number of authors (see among
others \cite{Fri00}, \cite{ManSch05}, \cite{MusZar04} and
\cite{HedHob08} for an overview). We should highlight at this
point that the indifference pricing mechanism is
\textit{subjective}, in the sense that a utility maximizer quotes
prices at which she is willing to buy or sell a given contingent
claim. However, there is no guarantee that these prices are the
ones at which any kind of transaction actually takes place.
Therefore, throughout this paper we prefer to call these prices
\textit{values} to emphasize their subjective nature.

One of the main flaws of the utility maximization (and of the
induced indifference valuation) is the dependence on the time
horizon at which the utility function stands. Although for
investment goals and single claim pricing, fixing a certain
investment/pricing time horizon may not be problematic, it creates
consistency concerns. In particular, this theory does not provide
a way to set another time horizon and the continuation of the
investments to be consistent. Similarly, the valuing of continent
claims with maturity later than the chosen time horizon can not be
addressed with the available tools. This is because there is no
forward shifting of a utility function. Even more inconvenient is
that fixing a utility at some time in the future leaves no room
for updating the utility function (and by extension the investment
goals) until the terminal horizon. It looks like an agent is stuck
with her utility function and a given subjective probability
measure, no matter what happens to the market, her endowment or
her beliefs.

The problem of time horizon dependence of investment choices has
been recently studied by a number of authors (\cite{ChiStrLi07},
\cite{ElKarMrad10}, \cite{HedHob07}, \cite{Zar09} and
\cite{ZarZit10}). A common concept of these works is that agents
aim to maximize, instead of a utility function, a family of
state-dependent utility functions in a time-consistent way. In
this paper, we work on the notion of \textit{forward performance}
or \textit{forward utility}, which has been introduced in the
works of M.~Musiela and T.~Zariphopoulou \cite{MusZar07} and
\cite{MusZar08} (see also \cite{Zar09} for an overview). In words,
this concept suggests that in contrast to the backward utility
function maximization, the agents choose a family of
state-dependent utility functions and their investment goal is to
find the admissible trading strategy that keeps the expectation of
this family at the same level (see Definition
\ref{def:forwardutility} for the exact definition and
\cite{MusZar10} for an extended discussion). If the family of
utility functions is of certain exponential type, the forward
performance is called \textit{exponential}. Explicit formulas for
the optimal strategy and the optimal wealth process under this
type of forward performance criteria has been provided in
\cite{MusZar08} under a Markovian market model (see also
\cite{ZarZit10} for some related
 discussion). More recently, G.~\v{Z}itkovi\'{c} in \cite{Zit09}
 establishes the characterization of the forward exponential
 performance process in a general semimartingale market model and as a
 special case in a diffusion stochastic volatility model, similar
 to the one we shall impose in the present paper. One of the important
 part of the characterization of the forward exponential utility
 functions is that the risk aversion becomes stochastic process
 instead of constant (as in the classic exponential utility
 function). Furthermore, for the stochastic risk aversion process,
 usually denoted by $\gamma_t$, it holds that the quantity
 $1/\gamma_t$, which can be thought as agent's (stochastic) risk
 tolerance, is replicable.

The first aim of this paper (Section 2) is to contribute to the
theory of exponential forward performance by investigating how
agents value contingent claims under such investment criteria.
Valuation in a forward manner has several differences in
comparison with the backward valuation (which is induced by
classic utility functions) both in financial and technical sense.
In Section 2, we state and prove a number of properties of the
forward indifference valuation and we point out the differences to
the corresponding backward valuation. Namely, based on the
characterization results in \cite{Zit09}, we are able to prove
that the dynamic version of the forward indifference valuation
solves a certain type of backward stochastic differential equation
(see Proposition \ref{pro: BSDE represantation}). This equation is
similar to the corresponding equation solved by the dynamic
exponential utility indifference price (provided in
\cite{ManSch05}), where the differences are limited to the
referred martingale probability measure and to that the risk
aversion becomes stochastic in the forward valuation. Using this
stochastic differential equation, we also show exactly how the
forward performance changes when the agent buys or sells a
contingent claim. The next step in analyzing the forward
exponential indifference valuation is its \textit{robust
representation}. As pricing functional in incomplete markets, the
forward indifference value can be seen as a convex map, i.e.~a
convex risk measure in the sense of \cite{FolSch02}. Following the
related literature on dynamic convex risk measure (see
\cite{DetSca05} and \cite{KloSch07}), we write the forward
indifference value as a minimum (over martingale probability
measures) of the expectation of the claim plus a penalty function,
which incorporates the agent's specific characteristics. This
representation is useful for proving several properties of the
indifference valuation such as continuity, differentiability with
respect to the units of the claim and monotonicity with respect to
risk aversion (see Propositions \ref{pro:price properties} and
\ref{pro:derivative}).

The second part of the paper (Section 3) is dedicated to the
\textit{optimal risk sharing} between two agents whose investment
criteria are based on forward exponential performances. Optimal
risk sharing problem is about two agents who negotiate the sharing
of their endowments in such a way that the sum of their positions
is optimized in terms of risk. This problem is well studied under
several models from classic utility functions (see \cite{Bor62}
and \cite{BuhJew79}) to convex risk measures (see among others
\cite{BarElk04}, \cite{BarElk05}, \cite{FilSvi08} and
\cite{JouSchTou08}). All of these models are set in a backward
fashion, that is the optimization criteria have a fixed time
horizon, which in fact equals to the maturity of the agents'
endowments. In this paper, we initiate this problem in the forward
setting and establish its solution in three different cases
regarding the model parameters: (a) when both agents have constant
risk aversions, (b) when agents have common but stochastic risk
aversions and (c) when agents have different and stochastic risk
aversions. In case (a), the problem is reduced to the backward
setting. As it is pointed out in Remark \ref{rem:constant gamma},
given a fixed time horizon, the forward exponential performance
criterion with constant risk aversion can be considered as a
simple utility function under a specific random endowment. By
exploiting the robust representation of the indifference value, we
get the exact form of the contracts that optimally share the
agents' random endowments and we bring out its similarity with the
classic entropic risk measure case (studied among others in
\cite{BarElk05}). In case (b), a similar closed form solution of
the optimal contract can be provided, where the impact of agents'
characteristics is clear (for example, the more different the
agents' beliefs about the probability measure are, the bigger the
expected size of the optimal contract is). Finally, for the more
general case of different and stochastic risk aversions we need to
look at the time evolution of the inf-convolution risk measure
induced by agents' forward performance criteria. We first
establish the necessary and sufficient conditions under which this
measure can be seen as one induced by some other forward
exponential performance. Then, we generalize the results of
\cite{BarElk05} in the forward setting and get the stochastic
differential equation satisfied by the inf-convolution measure.
This result enables us to derive the form of the optimal risk
sharing contract, which also allows the comparison with the
analogous backward valuation setting. The structure of the optimal
risk sharing contract admits a mild generalization in the case of
forward exponential performance. In particular, in the forward
setting the optimal risk sharing consists of three terms, one that
has to do with the sharing of the endowments, one that
incorporates the difference of beliefs between agents and a
replicable term (which can be ignored since it does not transfer
any risk). The first term has the same form as in the case of
entropic risk measures (however the risk aversion becomes
stochastic); the second term, i.e.~the sharing of agents'
difference of beliefs does not depend on agents' random
endowments, but does depend on agents' risk aversion levels.

The market model used in this manuscript consists of one riskless
asset and one risky asset, whose volatility is driven by a
generalized It\^{o} process (the same model has also been used in
\cite{SirZar05} and \cite{Zit09}). It should be pointed out that
the majority of the results in this paper can be generalized in a
straightforward way to models with more risky assets. We choose to
work in this simplified model in order to focus on the
interpretation of the results and the explanation of how agents
evaluate claims and share risks under forward looking investment
criteria. Furthermore, this work deals with the exponential type
of forward performance, since this type is more tractable and
closed form solutions can be provided. It also helps the
comparison with the backward case, where there are several
well-known results regarding valuation and risk sharing issues.
Finally, as mentioned and illustrated in \cite{MusZar08},
exponential forward performance is quite general and captures a
variety of agents' distinct characteristics.

\bigskip

\section{Market Setting}

\subsection{Assets and admissible strategies}

The market consists of a risky and a risk-free asset. The
risk-free asset is used as a num\'{e}raire and its price process
evolution is given by
$$dB_t=rB_tdt$$
where $r>0$ is a constant. The price process of the risky asset
satisfies the following stochastic differential equation (SDE)
\begin{equation}\label{eq:stock price}
    dS_t=\mu(Y_t)S_tdt+\sigma(Y_t)S_tdW^1_t
\end{equation}
where $S_0>0$ and process $Y$ solves the equation
\begin{equation}\label{eq:Y_t}
    dY_t=b(Y_t)dt+a(Y_t)\left(\rho dW^1_t+\sqrt{1-\rho^2}dW^2_t\right)
\end{equation}
Pair $\left(W^1_t,W^2_t\right)_{t\in [0,\infty)}$ is a
2-dimensional standard Brownian motion defined on a probability
space $(\Omega,\FF,\mathbb{F}, \PP)$, where
$\mathbb{F}=(\FF_t)_{t\geq 0}$ is the augmented $\sigma$-algebra
generated by $\left(W^1_t,W^2_t\right)_{t\in [0,\infty)}$. We
furthermore impose that $\sigma>0$ and $\rho$ is a constant
belonging in $(-1,1)$. As usual, we assume that the system of
equations \eqref{eq:stock price} and \eqref{eq:Y_t} has a unique
strong solution.

We define the \textit{market price of risk process}
$(\lambda_t)_{t\in[0,\infty)}$, via
\begin{equation}\label{eq:sharp ratio}
    \lambda_t=\lambda(Y_t)=\frac{\mu (Y_t)-r}{\sigma
    (Y_t)},\,\,\,\,\,t\in[0,\infty).
\end{equation}

Throughout this paper, we impose the following technical
assumption.
\begin{assumption}\label{ass:1}
For every $T> 0$, there exists $\varepsilon>0$ such that
$\EE\left[e^{(1/2+\varepsilon)\int_0^T\lambda_u^2du}\right]<\infty$.
\end{assumption}
We then define the set of admissible strategies
$$\AA=\{\mathbb{F}\text{-progressively measurable } \pi\, : \, \EE\left[\int_0^t\sigma^2(Y_s)\pi_s^2ds\right]<\infty, \,\,\forall t>0\}.$$

The discounted wealth process of an admissible strategy $\pi$ with
initial capital $x$ is denoted by $X^{x,\pi}$ and satisfies the
following SDE
\begin{equation}\label{eq:wealth process}
    dX^{x,\pi}_t=\sigma(Y_t)\pi_t(\lambda(Y_t)dt+dW_t^1)
\end{equation}
where $x$ is the initial wealth (when the initial wealth is zero,
we simply write $X^{\pi}$). We also define the set
$\AA^{\infty}=\{\pi\in\AA:\,\, X_t^{\pi}\in\linf(\FF_t), \forall
t\geq 0\}$.

For the model at hand, we introduce the following notations for
any time horizon $T>0$.
\begin{equation}\label{set:progressive T}
    \PPP_T=\{\text{$\mathbb{F}$-progressively measurable }\nu:\int_0^{T}\nu_u^2du<\infty, \text{ a.s.}\}
\end{equation}
\begin{equation}\label{set:progressive}
    \PPP=\underset{T>0}{\bigcap}\PPP_T
\end{equation}
and
\begin{equation}\label{set:progressive}
    \NN=\{(\beta,\nu)\in\PPP\times\PPP:Z^{\beta,\nu} \text{ is a
    true }\PP
    \text{-martingale}\},
\end{equation}
where $\left(Z^{\beta,\nu}\right)_{t\in[0,\infty)}$ is the
solution of the equation
$$dZ^{\beta,\nu}_u=-Z^{\beta,\nu}_u(\beta_udW_u^1+\nu_udW_u^2).$$
Note that under Assumption \ref{ass:1}, $(\lambda,0)\in\NN$.

For every arbitrarily chosen time horizon $T$ and every
$(\beta,\nu)\in\NN$, we define the probability measure
$\QQ^{\beta,\nu}\sim\PP_{|\FF_T}$ by its R-N derivative
$\frac{d\QQ^{\beta,\nu}}{d\left(\PP_{|\FF_T}\right)}=Z^{\beta,\nu}_T$.
We also define the set $\PPP^{\lambda}\subseteq\PPP$, which
contains all processes $\nu\in\PPP$ such that
$(\lambda,\nu)\in\NN$ (similarly we define the set
$\PPP_T^{\lambda}$).

A simple application of Girsanov Theorem implies that for every
$\nu\in\PPP^{\lambda}$ the discounted stock price
$\frac{S_t}{B_t}$ is a local-martingale under the measure
$\QQ^{\lambda,\nu}$. In fact, for the set of equivalent
local-martingale measures $\MM_T^e=\left\{\QQ\sim\PP:
\frac{S_t}{B_t} \text{ is a }\QQ\text{-local martingale in }
[0,T]\right\}$ it holds that:
$$\MM_T^e=\left\{\QQ^{\lambda,\nu}:
\nu\in\PPP_T^{\lambda}\right\}$$ (see \cite{ElKarQue95} for the
proof).

\subsection{The forward exponential performance criteria}

In this manuscript we assume that agent's investment goals are
modelled by so-called forward performance criteria (also called
forward or stochastic utilities and self-generating random
utilities) introduced in \cite{MusZar07} (see also \cite{MusZar09}
and \cite{Zit09}).

\begin{definition}\label{def:forwardutility}
A map $U:\Omega\times[0,\infty)\times\R\longmapsto\R$ is called a
\textit{forward performance process} if:
\begin{itemize}
    \item [(i)] It is measurable with respect to the product of
    the optional $\sigma$-algebra on $\Omega\times[0,\infty)$ and
    the Borel $\sigma$-algebra on $\R$.
    \item [(ii)] For fixed $\omega\in\Omega$ and $t\in[0,\infty)$, the mapping $x\longmapsto
    U_t(x)$ is strictly increasing and strictly concave.
    \item [(iii)] For all $s\geq t$ and $X\in\linf(\FF_t)$
\begin{equation}\label{eq: inequality def}
U_t(X)=\underset{\pi\in\AA^{\infty}}{\esssup}\EE\left[U_s(X_s^{X,\pi})|\FF_t\right]
\end{equation}
\end{itemize}
A forward performance is called \textit{exponential} if there
exist processes $(A_t)_{t\in[0,\infty)}$ and
$(\gamma_t)_{t\in[0,\infty)}$ such that $\gamma_t>0$ a.s. for
every $t\geq 0$ and
\begin{equation}\label{eq:exponential utility}
    U_t(x)=-e^{-\gamma_tx+A_t}
\end{equation}
for $x\in\R$, and $t\geq 0$.
\end{definition}
\begin{remark}
We follow the definition given in \cite{Zit09}, which does not
require that the optimal strategy is attained. The reason for this
choice it that we prefer to focus on pricing and risk sharing
issues rather than the technicalities on the existence of the
optimal strategy (see also the related discussion in
\cite{Zit09}.) An analysis of the portfolio management problem
with forward exponential criteria has been provided in
\cite{MusZar08}, where the authors give explicit formulas for the
optimal portfolio $\pi^*$ and the associated optimal wealth
process $X^{x,\pi^*}$, for a variety of model parameters.
\end{remark}

When agent's risk preferences are modelled by a utility function
$U(x)$, her investment criterion (up to some certain time horizon
$T$) is the maximization of the expected utility function. That is
for every $t\in[0,T]$, optimal trading strategy is defined by
\begin{equation}\label{eq:utility maximization}
\underset{\pi\in\AA}{\esssup}\EE[U(X_T^{x,\pi})|\FF_t].
\end{equation}
The utility given by $U(x)=-e^{-\gamma x}$ is called
\textit{exponential}.

The forward performance process can be seen as a modification of
the above utility maximization investment criterion, in the
following sense: There is no terminal time horizon set and the
choice of the utility is made at time 0,
i.e.~$U_0(x)=-e^{\gamma_0x+A_0}$. For every future time $t$, the
utility is updated by replacing the risk aversion coefficient
$\gamma_0$ with a stochastic one $\gamma_t$, and the term $A_0$
with $A_t$. Hence, we shall call the investment criterion
\eqref{eq:utility maximization} \textit{backward} to emphasize its
main difference with the forward performance criteria.

The following characterization of the forward exponential
performance processes has been proven in \cite{Zit09}.

\begin{theorem}[\v{Z}itkovi\'{c}, 2009]\label{thm:Gordan}
Suppose that Assumption \ref{ass:1} holds and let
$U_t(x)=-e^{-\gamma_tx+A_t}$ be a forward exponential performance
where $\frac{1}{\gamma_t}\in\linf(\FF_t)$ for all $t\geq 0$. Then,
there exist processes $(\vartheta_t)_{t\in[0,\infty)}$,
$(\phi_t)_{t\in[0,\infty)}$ and $(p_t)_{t\in[0,\infty)}$ such that
$\vartheta\in\AA^{\infty}$ and
$\frac{1}{\gamma_t}=X_t^{\frac{1}{\gamma_0},\vartheta}$, with
$\gamma_0\in\R_+$, $\phi,p\in\PPP$ and
\begin{equation}\label{eq:At characterization}
    A_t=A_0+\frac{1}{2}\int^t_0\left(\lambda(Y_u)-\delta_u\right)^2du +\gamma_tX^{p}_t-\frac{1}{2}\int^t_0\phi^2_udu-
    \int^t_0\phi_udW^2_u
\end{equation}
where $\delta_t=\gamma_t\vartheta_t\sigma(Y_t)$.
\end{theorem}

\begin{example}
A simple example of a forward exponential performance is the case
where $\vartheta_t=\phi_t=0$ for all $t\geq 0$. This corresponds
to constant risk aversion coefficient ($\gamma$ is not a
stochastic process). If we further set
$p_t=-\frac{\lambda(Y_t)}{\gamma\sigma(Y_t)}$, for every time
$t\geq 0$, we have that
$A_t=A_0-\frac{1}{2}\int_0^t\lambda^2(Y_u)du-\int_0^t\lambda(Y_u)dW_u^1$
(an example that has been used in \cite{Zar09}).
\end{example}

\begin{remark}
In view of the characterization of the forward exponential
performance, the process $p$ can be omitted when it belongs in
$\AA^{\infty}$. Similarly and without loss of generality we may
ignore the initial term $A_0$. Note also that the boundedness
assumption of $1/\gamma_t$ is not very restrictive in financial
sense, since for a given time $t$, $1/\gamma_t$ denotes agent's
the risk tolerance, which is normally a bounded quantity.
\end{remark}

Assuming that $p\in\AA^{\infty}$, the characterization of forward
exponential performances in Theorem \ref{thm:Gordan} allows us to
identify the exact elements of these performance criteria. More
precisely, the decomposition of process $A_t$ consists of two
parts:
\begin{itemize}
    \item The integral
    $\int^t_0\left(\lambda(Y_u)-\delta_u\right)^2du$,
which reflects how the agent incorporates the market's development
in her investment criteria, in a way that this incorporation takes
into account her stochastic risk tolerance level.
    \item The sum $\frac{1}{2}\int^t_0\phi^2_udu+
    \int^t_0\phi_udW^2_u$. This term does not depend on the level of
    risk aversion $\gamma_t$ and it can be considered as the way the changes of the
    unhedgeable source of the market make the agent
    update her subjective probability measure $\PP$ (see also the related comment in
    \cite{ZarZit10}). This is because $-e^{-\gamma_tx+A_t}$ can be
    written as
    $-e^{-\gamma_tx+\frac{1}{2}\int^t_0\left(\lambda(Y_u)-\delta_u\right)^2du}Z_t^{0,\phi}$.
\end{itemize}

In what follows we will identify a forward exponential performance
process (for which $p\in\AA^{\infty}$) by its
\textit{characterization pair}
$(\vartheta_t,\phi_t)_{t\in[0,\infty)}$, where
$\vartheta\in\AA^{\infty}$ and $\phi\in\PPP$ for all $t\geq 0$.
The associated process $A$ will be called the
\textit{characteristic process} of the forward exponential
performance.

An important difference between the forward and the standard
(backward) exponential investment criteria (defined in
\eqref{eq:utility maximization}) is that in the former when the
the optimal strategy in \eqref{eq: inequality def} is attained, it
\textit{does} depend on the initial wealth. However, this
dependence is quite clear (thanks to the replicability of
$1/\gamma_t$).

\begin{proposition}\label{pro:depedence}
Let $(\vartheta_t,\phi_t)_{t\in[0,\infty)}$ be the
characterization pair of a forward exponential performance. If the
optimal trading strategy process in \eqref{eq: inequality def} is
attained, then
\begin{equation}\label{eq:strategy and in.wealth}
    \pi_t^*(y)=\pi_t^*(x)+\gamma_0(y-x)\vartheta_t
\end{equation}
and
\begin{equation}\label{eq:wealth and in.wealth}
    X^{y,\pi^*(y)}_t-X^{x,\pi^*(x)}_t=\frac{\gamma_0}{\gamma_t}(y-x).
\end{equation}
for all $x,y\in\R$ and $t\geq 0$, where $\pi^*(x)$ stands for the
optimal strategy with initial wealth $x$.
\end{proposition}

\begin{proof}
The proof is a straightforward consequence of the facts that
$\frac{1}{\gamma_t}$ is replicable and that the backward
exponential indifference valuation is independent on the initial
wealth (see for instance \cite{HedHob08}).
\end{proof}

In the rest of this manuscript, we assume that the agents' initial
wealth is equal to zero. For any nonzero initial wealth we may
apply Proposition \ref{pro:depedence}.

\bigskip

\section{Valuation of Contingent Claims based on Forward Indifference}

If an agent's investment goals are determined by a forward
exponential performance, it is reasonable to suppose that she uses
indifference arguments in order to give values to contingent
claims. The idea of indifference valuation was introduced in the
finance literature in \cite{HodNeu89} and then deve\-loped and
analyzed for a number of utility functions and market settings
(see among others \cite{HedHob08} and the references therein).
This (subjective) valuation concept compares two situations, the
one where a contingent claim is bought or sold and another where
there is no transaction on this claim.

For the model at hand, for a certain time horizon $T>0$ we
consider an $\FF_T$-measurable payoff $C$. The (buyer)
indifference value is the price $p$ that makes the agent
indifferent between buying the claim at $p$ and not buying it at
all. In our forward performance setting the buyer's value of a
payoff $C$ at any time $t\in[0,T]$, denoted by $v^{(b)}_t(C)$ is
the $\FF_t$-measurable solution of the following equation
\begin{equation}\label{eq:indifference pricea}
        U_t\left(x\right)=
    \underset{\pi\in\AA^{\infty}}{\esssup}\EE\left[\left. U_T\left(x-v^{(b)}_t(C)+C+\int_t^T\pi_sdS_s\right)\right\vert\FF_t\right]
\end{equation}
that is
\begin{equation}\label{eq:indifference priceb}
        -e^{-\gamma_tx+A_t}=
    \underset{\pi\in\AA^{\infty}}{\esssup}\EE\left[\left. -e^{-\gamma_T\left(x-v^{(b)}_t(C)+C+\int_t^T\pi_sdS_s\right)+A_T} \right\vert\FF_t\right]
\end{equation}
Due to replicability of $1/\gamma_t$, \eqref{eq:indifference
priceb} can equivalently be written as
\begin{equation}\label{eq:indifference price}
        -e^{-\gamma_t(x+v^{(b)}_t(C))+A_t}=
        \underset{\pi\in\AA^{\infty}}{\esssup}\EE\left[\left. -e^{-\gamma_T\left(x+C+\int_t^T\pi_sdS_s\right)+A_T} \right\vert\FF_t\right]
\end{equation}
Hence we get that the forward indifference value at time $t$ of a
claim $C$ is given by
$$v^{(b)}_t(C)=
    -x-\frac{A_t}{\gamma_t}+\frac{1}{\gamma_t}\log\left(\underset{\pi\in\AA^{\infty}}
    {\essinf}\EE\left[\left. e^{-\gamma_T\left(x+C+\int_t^T\pi_sdS_s\right)+A_T} \right\vert\FF_t\right]\right)$$
and again by the replicability of $1/\gamma_t$ we have that
\begin{equation}\label{eq:indifference price formula}
        v^{(b)}_t(C)=
    -\frac{A_t}{\gamma_t}+\frac{1}{\gamma_t}\log\left(\underset{\pi\in\AA^{\infty}}
    {\essinf}\EE\left[\left. e^{-\gamma_T\left(C+\int_t^T\pi_sdS_s\right)+A_T} \right\vert\FF_t\right]\right)
\end{equation}
Note that that the indifference value $v^{(b)}_t(C)$ does not
depend on the initial wealth (see also \cite{LeuSirZar10},
\cite{MusSokZar10}, \cite{MusZar04} and \cite{ZarZit10}).

Results on the value that solves equation \eqref{eq:indifference
pricea} have been provided in \cite{MusZar04} and in
\cite{MusSokZar10} for specific types of forward exponential
performances in a stochastic factor model and in a binomial-type
market model respectively. The use of forward exponential
performance criteria for pricing contingent claims of
American-type has been studied in \cite{LeuSirZar10} under the
assumption of constant risk aversion.

One interesting question about these pricing mechanisms is how the
forward exponential indifference valuation differs with the
indifference valuation induced by exponential utilities. More
precisely, we want to compare the solution $v^{(b)}_t(C)$ of
\eqref{eq:indifference pricea}, with the solution $pr^{(b)}_t(C)$
of the following equation
\begin{equation}\label{eq:back indifference price}
        \underset{\pi\in\AA^{\infty}}{\esssup}\EE\left[\left. -e^{-\gamma(x+pr^{(b)}_t(C)+\int_t^T\pi_sdS_s)}\right\vert\FF_t\right] =
    \underset{\pi\in\AA^{\infty}}{\esssup}\EE\left[\left. -e^{-\gamma(x+C+\int_t^T\pi_sdS_s)}\right\vert\FF_t\right]
\end{equation}
where $\gamma\in\R_+$ is the risk aversion coefficient. Throughout
this paper the price process $pr^{(b)}(C)$ will be called
\textit{backward exponential indifference value process}.

Such a comparison has been studied in \cite{MusSokZar10} and
\cite{MusZar07} (for constant risk aversion process). In the
present section, we aim to extend the results on forward
indifference valuation for general forward exponential performance
process and by doing so to highlight the special properties of the
forward valuation regarding its comparison with the backward
exponential valuation.

\subsection{The BSDE representation of the indifference value}
It has been proved in \cite{ManSch05} that, under continuous
filtration, the indifference value process under backward
exponential utility sa\-tisfies a certain type of backward
stochastic differential equation (BSDE). Adapted to our market
model, the (buyer) indifference value of a contingent claim
$C\in\linf(\FF_T)$ satisfies the following BSDE
\begin{equation}\label{eq:BSDE for backward}
    C_t=C_T-\frac{\gamma}{2}\int_t^T\zeta^2_udu-\int_t^T\zeta_ud\hat{W}^2_u-\int_t^T\theta_udS_u
\end{equation}
and \begin{equation}\label{eq:terminal condition} C_T=C,
\end{equation}
where $\gamma\in\R_+$ is the constant risk aversion coefficient,
$\theta\in\AA$, $(\hat{W}^2_t)_{t\in[0,T]}$ is a Brownian motion
under minimal entropy martingale measure and under this measure
$\left(\int_0^{t}\zeta_udW^2_u\right)_{t\in[0,T]}$ is a true
martingale. We call the triple
$(C_t,\zeta_t,\theta_t)_{t\in[0,T]}$ a solution of BSDE
\eqref{eq:BSDE for backward} with terminal condition
\eqref{eq:terminal condition}. Theorem 13 in \cite{ManSch05}
guarantees that for $C\in\linf(\FF_T)$ there is a unique uniformly
bounded solution.

The following proposition establishes that the above
representation has a nice extension in the case of the forward
exponential performance.

\begin{proposition}\label{pro: BSDE represantation}
Let $(\vartheta_t, \phi_t)_{t\in[0,\infty)}$ be the
characterization pair of a forward exponential performance and
assume that there exists a constant $K_{\gamma}$ such that
$\underset{t\in[0,T]}{\sup}||\gamma_t||_{\linf}\leq K_{\gamma}$
and that $\phi\in\PPP^{\lambda}_T$. The forward exponential
indifference (buyer) value process of a contingent claim
$C\in\linf(\FF_T)$ is the unique uniformly bounded solution,
$(C_t)_{t\in[0,T]}$, of the following BSDE under the martingale
measure $\QQ^{\lambda, \phi}$
\begin{equation}\label{eq: BSDE of price}
    C_t=C_T-\frac{1}{2}\int_t^T\gamma_u\zeta^2_udu-\int_t^T\zeta_udW^{2,\phi}_u-\int_t^T\theta_udS_u
\end{equation}
and
\begin{equation}\label{eq: boundary condition}
    C_T=C
\end{equation}
for some processes $(\theta_t,\zeta_t)_{t\in[0,\infty)}$, such
that
$$\EE_{\QQ^{\lambda,\phi}}\left[\int_0^T\theta_u^2du\right]<\infty\,\,\,\text{and
}\,\,\,
\EE_{\QQ^{\lambda,\phi}}\left[\int_0^T\zeta_u^2du\right]<\infty,$$
where $W^{2,\phi}_t=W_t^2+\int^t_0\phi_udu$.
\end{proposition}

\begin{proof}
Recall the indifference valuation problem \eqref{eq:indifference
price}
\begin{equation}\label{eq:exponential indifference price}
-e^{-\gamma_tv^{(b)}_t(C)+A_t}=
    \underset{\pi\in\AA^{\infty}}{\esssup}\EE\left[\left. -e^{-\gamma_T\int_t^T\pi_sdS_s-\gamma_TC+A_T}\right\vert\FF_t\right]
\end{equation}
where process $A$ is given by the characterization \eqref{eq:At
characterization}, with $A_0=0$. We now define the process
$(\tilde{A}_t)_{t\in[0,\infty)}$ as
\begin{equation}\label{eq:new phi}
    \tilde{A}_t=\left\{
\begin{array}{ll}
    -\gamma_tv^{(b)}_t(C)+A_t, & \hbox{$t\leq T$;} \\
    -\gamma_tC+A_t, & \hbox{$t>T$.} \\
\end{array}%
\right.
\end{equation}
By the replicability of $1/\gamma_t$ we get that for every $t\geq
0$
\begin{equation}\label{eq:proof1}
-e^{-\gamma_tX+\tilde{A}_t}=
    \underset{\pi\in\AA^{\infty}}{\esssup}\EE\left[\left. -e^{-\gamma_TX -\gamma_T\int_t^T\pi_sdS_s+\tilde{A}_T}\right\vert\FF_t\right]
\end{equation}
for every $X\in\linf(\FF_t)$. In the view of Definition
\ref{def:forwardutility}, problem \eqref{eq:proof1} leads to
another forward exponential performance, where the risk aversion
process remains $\gamma_t$ and the characteristic process is given
by $\tilde{A}_t$. By Theorem \ref{thm:Gordan}, there exist
processes $z,\tilde{p}\in\PPP$ such that
$$\tilde{A}_t=\tilde{A}_0+\frac{1}{2}\int^t_0\left(\lambda(Y_u)-\delta_u\right)^2du +\gamma_tX^{\tilde{p}}_t-\frac{1}{2}\int^t_0z^2_udu-
    \int^t_0z_udW^2_u.$$
Hence, for any $t\in[0,T]$
\begin{eqnarray*}
  v^{(b)}_t(C) &=& \frac{A_t-\tilde{A}_t}{\gamma_t} \\
    &=& \frac{1}{\gamma_t}\left(A_0-\tilde{A}_0+\gamma_tX^{\tilde{p}}_t-
    \frac{1}{2}\int^t_0(\phi_u^2-z_u^2)du-\int^t_0(\phi_u-z_u)dW^2_u\right) \\
   &=& -\frac{\tilde{A}_0}{\gamma_0} + X_t^{\tilde{p}-A_0\vartheta}-\frac{1}{\gamma_t}\left(
    \frac{1}{2}\int^t_0(\phi_u^2-z_u^2)du+\int^t_0(\phi_u-z_u)dW^2_u\right).
\end{eqnarray*}
Note that with the above notation
$v^{(b)}_0(C)=-\frac{\tilde{A}_0}{\gamma_0}$ and $W^{2,\phi}$ is a
Brownian motion under the measure $\QQ^{\lambda, \phi}$, strongly
orthogonal to $S$.

Hence, the indifference value process satisfies the following
equation
$$v^{(b)}_t(C)=v^{(b)}_0(C)+X_t^{\tilde{p}}+\frac{1}{\gamma_t}\left(
    \frac{1}{2}\int^t_0(\phi_u-z_u)^2du+\int^t_0(\phi_u-z_u)dW^{2,\phi}_u\right).$$
Let
$\Lambda_t=\frac{1}{2}\int^t_0(\phi_u-z_u)^2du+\int^t_0(\phi_u-z_u)dW^{2,\phi}_u$,
for every $t\in [0,\infty)$. A simple application of Ito's formula
implies that
$$\frac{\Lambda_t}{\gamma_t}= \frac{1}{2}\int^t_0\frac{(\phi_u-z_u)^2}{\gamma_u}du+\int^t_0\frac{(\phi_u-z_u)}{\gamma_u}dW^{2,\phi}_u+
\int_0^t\Lambda_u\vartheta_u\sigma(Y_u)dW^{1,\lambda}_u,$$ where
$W^{1,\lambda}_t=W_t^1+\int^t_0\lambda_udu$. Note that the process
$\zeta=\frac{\phi-z}{\gamma}$ belongs in $\PPP_T$, thanks to
uniform boundedness of $\gamma_t$ for all $t\in[0,T]$ and the fact
that $\phi,z\in\PPP_T$. It is left to set
$\theta_t=(\Lambda_t\vartheta_u+\tilde{p}_t)\sigma(Y_t)\in\PPP$
and apply Lemma \ref{lem:bounds} below to get the integrability of
$\int_0^T\theta_u^2du$ and $\int_0^T\zeta_u^2du$ under the measure
$\QQ^{\lambda,\phi}$.

Finally, the uniqueness of the solution follows from Proposition
\ref{pro:comparison} for $\gone=\gtwo$ and $\phione=\phitwo$.
\end{proof}

\begin{lemma}\label{lem:bounds}
Impose the same assumptions as in Proposition \ref{pro: BSDE
represantation} and let $(C_t, \zeta_t,\theta_t)$ be the solution
of \eqref{eq: BSDE of price} and terminal condition \eqref{eq:
boundary condition} for some contingent claim $C\in\linf(\FF_T)$,
where $(C_t)_{t\in[0,T]}$ is uniformly bounded. Then, there exists
a constant $K>0$ such that
\begin{center}
$\underset{\tau}{\sup}\EE_{\QQ^{\lambda,\phi}}\left[\left.\int_{\tau}^T\sigma^2(Y_t)\theta^2_tdt\right\vert\FF_{\tau}\right]+
\underset{\tau}{\sup}\EE_{\QQ^{\lambda,\phi}}\left[\left.\int_{\tau}^T\zeta_t^2dt\right\vert\FF_{\tau}\right]<K$
\end{center}
where the supremum is taken under any stopping time
$\tau\in[0,T]$.
\end{lemma}

\begin{proof}
We first apply the It\^{o}'s formula for the process $e^{C_t}$
which implies that
\begin{equation}\label{eq:exp(Ct)}
    d(e^{C_t})=e^{C_t}\left(\frac{\zeta_t^2}{2}+\frac{\gamma_t\zeta_t^2}{2}+\frac{\theta_t^2\sigma^2(Y_t)}{2}\right)dt+e^{C_t}\zeta_td\tilde{W}^2_t+e^{C_t}\theta_t\sigma(Y_t)d\tilde{W}_t^1
\end{equation}
where, $W^{1,\lambda}_t=W_t^1+\int^t_0\lambda_udu$. There is a
sequence of stopping times $\tau_n$, with $\tau_n\nearrow T$ such
that $\int_0^{t\wedge\tau_n}\zeta_udW^{2,\phi}_u$ and
$\int_0^{t\wedge\tau_n}\theta_u\sigma_udW^{1,\lambda}_u$ are
$\QQ^{\lambda,\phi}$-martingales. The boundness of $C$ implies
that for any stopping time $\tau\in[0,T]$
$$e^{||C||_{\infty}}\geq\EE_{\QQ^{\lambda,\phi}}\left[\left.e^{C_{\tau_n}}-e^{C_{\tau\wedge\tau_n}}\right\vert\FF_{\tau\wedge\tau_n}\right]=
\EE_{\QQ^{\lambda,\phi}}\left[\left.\int_{\tau\wedge\tau_n}^{\tau_n}e^{C_t}\left(\frac{\zeta_t^2+\sigma^2(Y_t)\theta^2_t+\gamma_t\zeta^2_t}{2}\right)dt\right\vert\FF_{\tau\wedge\tau_n}\right]$$
$\,\,\,\,\,\,\,\,\,\,\,\,\,\,\,\,\,\,\,\,\,\,\,\,\,\,\,\geq\EE_{\QQ^{\lambda,\phi}}\left[\left.\int_{\tau_n}^{\tau\wedge\tau_n}e^{C_t}\left(\frac{\zeta_t^2+\sigma^2(Y_t)\theta^2_t}{2}\right)dt\right\vert\FF_{\tau\wedge\tau_n}\right]$
$$\geq
\frac{e^{-||C||_{\infty}}}{2}\EE_{\QQ^{\lambda,\phi}}\left[\left.\int_{\tau\wedge\tau_n}^{\tau_n}\zeta^2_tdt\right\vert\FF_{\tau\wedge\tau_n}\right]
+\frac{e^{-||C||_{\infty}}}{2}\EE_{\QQ^{\lambda,\phi}}\left[\left.\int_{\tau\wedge\tau_n}^{\tau_n}\sigma^2(Y_t)\theta^2_tdt\right\vert\FF_{\tau\wedge\tau_n}\right]$$

Hence,
$\EE_{\QQ^{\lambda,\phi}}\left[\left.\int_{\tau\wedge\tau_n}^{\tau_n}\zeta^2_tdt\right\vert\FF_{\tau\wedge\tau_n}\right]
+\EE_{\QQ^{\lambda,\phi}}\left[\left.\int_{\tau\wedge\tau_n}^{\tau_n}\sigma^2(Y_t)\theta^2_tdt\right\vert\FF_{\tau\wedge\tau_n}\right]\leq
2e^{||C||_{\infty}}$. Letting $n\rightarrow\infty$ completes the
proof.
\end{proof}

\begin{corollary}\label{cor:new phi}
Impose the same assumptions as in Proposition \ref{pro: BSDE
represantation} and let $(C_t, \zeta_t,\theta_t)$ be the solution
of \eqref{eq: BSDE of price} and terminal condition \eqref{eq:
boundary condition} for some contingent claim $C\in\linf(\FF_T)$,
where $(C_t)_{t\in[0,T]}$ is uniformly bounded. If the agent sells
the claim at price $C_0$, then the characterization process of her
forward exponential performance criteria becomes
$$\tilde{A}_t=-\frac{C_0}{\gamma_0}+\frac{1}{2}\int^t_0\left(\lambda(Y_u)-\delta_u\right)^2du +\gamma_tX^{\theta}_t-\frac{1}{2}\int^t_0(\phi_u^C)^2du-
    \int^t_0\phi^C_udW^2_u$$
where
\begin{equation}\label{eq:new phi}
    \phi^C_t=\left\{
\begin{array}{ll}
    \phi_t-\gamma_t\zeta_t, & \hbox{$t\leq T$;} \\
    \phi_t, & \hbox{$t>T$,} \\
\end{array}%
\right.
\end{equation}
and $\theta$ a process for which $\theta_t=0$ for every $t>T$.
\end{corollary}

\begin{remark}
Note that the constant risk aversion of equation \eqref{eq:BSDE
for backward} becomes stochastic in \eqref{eq: BSDE of price} in a
mild manner. In addition, the minimal entropy martingale measure
is replaced by the measure $\QQ^{\lambda,\phi}$.
\end{remark}

\subsection{The robust representation}

As in the backward indifference valuation, the forward valuation
can be considered as a dynamic (convex) risk measure in the sense
of \cite{DetSca05} (see also \cite{ZarZit10}). For a fixed time
horizon $T$, the map $-v^{(b)}_t(\cdot):\linf(\FF_T)\longmapsto
\linf(\FF_t)$ is convex, cash invariant and decreasing. In the
following theorem, we state its robust representation. For this we
need to define the set of martingale measures with finite entropy
\begin{equation}\label{eq:finite entropy}
    \PPP_T^H:=\{\nu\in\PPP_T^{\lambda}:\, \EE_{\QQ^{\lambda,\nu}}[\log
    Z_T^{\lambda,\nu}]<\infty\}.
\end{equation}

\begin{theorem}\label{thm:robust repre}
Impose Assumption \ref{ass:1}, let
$(\gamma_t,\phi_t)_{t\in[0,\infty)}$ be the characterization of a
forward exponential performance and $T>0$ some time horizon. If we
assume that there exist constants $K_{\gamma}, \epsilon >0$ such
that
\begin{itemize}
    \item [(i)] $\EE[e^{(1+\epsilon)\int_0^T\phi^2_udu}]<\infty$
    and
    \item [(ii)]
    $\underset{t\in[0,T]}{\sup}||\gamma_t||_{\linf}<K_{\gamma}$,
\end{itemize}
the forward indifference (buyer) valuation process
$(v^{(b)}_t(\cdot))_{t\in[0,T]}$ defined in \eqref{eq:indifference
price} has the following representation
\begin{equation}\label{eq:robust repre}
v^{(b)}_t(C)=\underset{\nu\in\PPP_T^{H}}{\essinf}\{\EE_{\QQ^{\lambda,\nu}}[C|\FF_t]+\alpha_{t,T}(\QQ^{\lambda,\nu})\}
\end{equation}
for every $C\in\linf(\FF_T)$, where
\begin{equation}\label{eq:penalty}
    \alpha_{t,T}(\QQ^{\lambda,\nu})=\frac{1}{2}\EE_{\QQ^{\lambda,\nu}}\left[\left.
    \int_t^T\frac{(\nu_s-\phi_s)^2}{\gamma_s}\right\vert\FF_t\right].
\end{equation}
Furthermore, the infimum in \eqref{eq:robust repre} is attained by
the process $\nu^*_t=\phi_t+\zeta_t\gamma_t$, for $t\in[0,T]$,
where $\zeta$ is the corresponding part of the solution of BSDE
\eqref{eq: BSDE of price}.
\end{theorem}

\begin{proof}
We fix an arbitrary chosen contingent claim $C\in\linf(\FF_T)$
following the steps in the proof of Proposition \ref{pro: BSDE
represantation} we set $\tilde{A}_t=-\gamma_tv^{(b)}_t(C)+A_t$,
which is the characterization process of the forward exponential
utility after selling claim $C$. Furthermore, from Theorem 4.4 in
\cite{Zit09}, we have that
\begin{equation}\label{eq:proof of penalty1}
    \frac{1}{\gamma_t}\left(\log\left(\frac{1}{\gamma_t}\right)-1-\tilde{A}_t\right)=
     \underset{\QQ\in\MM^e_T}{\essinf}\EE_{\QQ}\left[\left.\frac{1}{\gamma_T}
     \left(\log\left(\frac{1}{\gamma_T}\frac{Z^{\QQ}_T}{Z^{\QQ}_t}\right)-1-\tilde{A}_T\right)\right|\FF_t\right]
\end{equation}
which implies that
\begin{equation}\label{eq:proof of penalty2}
    \frac{1}{\gamma_t}\left(\log\left(\frac{1}{\gamma_t}\right)-1-A_t\right)+v^{(b)}_t(C)=\underset{\QQ\in\MM^e_T}{\essinf}\EE_{\QQ}\left[\left.\frac{1}{\gamma_T}\left(\log\left(\frac{1}{\gamma_T}\frac{Z^{\QQ}_T}{Z^{\QQ}_t}\right)-1-A_T\right)
    +C\right|\FF_t\right].
\end{equation}
A simple rearrangement of the terms gives that indifference value
process has the following dual representation
\begin{equation}\label{eq:dual representation}
    v^{(b)}_t(C)=\underset{\QQ\in\MM^e_T}{\essinf}\{\EE_{\QQ}[C|\FF_t]+H_t(\QQ,T)\}-H^{(0)}_t(T).
\end{equation}
where $$H_t(\QQ,T)=\EE\left[\left.
h\left(\frac{1}{\gamma_T}\frac{Z^{\QQ}_T}{Z^{\QQ}_t}\right)-\frac{1}{\gamma_T}\frac{Z^{\QQ}_T}{Z^{\QQ}_t}A_T\right\vert\FF_t\right]$$
with $h(y)=y\log(y)-y$ for $y>0$, and
$H^{(0)}_t(T)=\underset{\QQ\in\MM^e_T}{\essinf}H_t(\QQ,T)$. We
first show that for every $\nu\in\PPP_T^H$
\begin{equation}\label{eq: proof1}
    H_t(\QQ^{\lambda,\nu},T)=h\left(\frac{1}{\gamma_t}\right)-\frac{A_t}{\gamma_t}+\frac{1}{2}\EE_{\QQ^{\lambda,\nu}}\left[\left.
    \int_t^T\frac{(\nu_s-\phi_s)^2}{\gamma_s}\right\vert\FF_t\right].
\end{equation}
Indeed,
\begin{eqnarray*}
H_t(\QQ^{\lambda,\nu},T) &=& \EE\left[\left.
h\left(\frac{1}{\gamma_T}\frac{Z^{\lambda, \nu}_T}{Z^{\lambda,
\nu}_t}\right)-\frac{1}{\gamma_T}\frac{Z^{\lambda,
\nu}_T}{Z_t^{\lambda, \nu}}A_T
   \right\vert\FF_t\right]= \EE\left[\left. h\left(\frac{1}{\gamma_t}\frac{Z^{\lambda-\delta, \nu}_T}{Z^{\lambda-\delta, \nu}_t}\right)-\frac{1}{\gamma_t}\frac{Z^{\lambda-\delta, \nu}_T}{Z_t^{\lambda-\delta, \nu}}A_T
   \right\vert\FF_t\right]\\
&=& \frac{1}{\gamma_t}\EE_{\QQ^{\lambda-\delta,
\nu}}\left[\left.\log\left(\frac{1}{\gamma_t}\frac{Z^{\lambda-\delta,
\nu}_T}{Z^{\lambda-\delta, \nu}_t}\right)-A_T
   \right\vert\FF_t\right]-\frac{1}{\gamma_t}\\
   &=&  h\left(\frac{1}{\gamma_t}\right)+\frac{1}{\gamma_t}
   \EE_{\QQ^{\lambda-\delta, \nu}}\left[\left.\log\left(\frac{Z^{\lambda-\delta, \nu}_T}{Z^{\lambda-\delta, \nu}_t}\right)-A_T
   \right\vert\FF_t\right]\\
&=&
h\left(\frac{1}{\gamma_t}\right)-\frac{A_t}{\gamma_t}+\frac{1}{2\gamma_t}\EE_{\QQ^{\lambda-\delta,
\nu}}
   \left[\left. \int_t^T(\nu_u-\phi_u)^2du\right\vert\FF_t\right]\\
   & + & \frac{1}{\gamma_t}\EE_{\QQ^{\lambda-\delta, \nu}}
   \left[\left.
   -\int_t^T(\lambda_u-\delta_u)dW^{1,\lambda-\delta}_u-\int_t^T(\nu_u-\phi_u)dW^{2,\nu}_u\right\vert\FF_t\right]
\end{eqnarray*}
where
$W^{1,\lambda-\delta}_t=W^1_t+\int_0^t(\lambda_u-\delta_u)du$ and
$W^{2,\nu}_t=W^2_t+\int_0^t\nu_udu$, for $t\in[0,T]$. Now, for
every $\nu\in\PPP_T^{H}$, the expectation of the stochastic
integrals is zero. For this, thanks to the uniform boundedness of
process $\gamma$, it is enough to show that
$$\EE_{\QQ^{\lambda, \nu}}\left[\int_0^T(\lambda_u-\delta_u)^2du+\int_0^T(\nu_u-\phi_u)^2du\right]<\infty.$$
By item (i) and the inequality $y\log y-y+e^x\leq xy$, which holds
for every $y>0$ and $x\in\R$, it follows that
$\int_0^T\lambda_u^2du$ and $\int_0^T\phi_u^2du$ belong in
$\lone(\QQ^{\lambda,\nu},\FF_T)$. Also, finite entropy of
$\QQ^{\lambda,\nu}$ implies that $\EE_{\QQ^{\lambda,
\nu}}\left[\int_0^T\nu_u^2du\right]<\infty$ (see the proof of
Theorem 1.9 in \cite{Kaz94}), whereas the same condition for
process $\delta$ is guaranteed by the uniform boundedness of
$\gamma$ (see also Lemma \ref{lem:bounds}). To finish the proof of
equation \eqref{eq: proof1}, it is left to observe that
replicability of $\left(\frac{1}{\gamma_t}\right)_{t\in[0,T]}$
implies that
$$\EE_{\QQ^{\lambda,\nu}}\left[\left.\frac{1}{\gamma_T}\int_t^T(\nu_s-\phi_s)^2\right\vert\FF_t\right]=
\EE_{\QQ^{\lambda,\nu}}\left[\left.\int_t^T\frac{(\nu_s-\phi_s)^2}{\gamma_s}\right\vert\FF_t\right].$$
From Proposition 4.9 in \cite{Zit09}, we have that for every
$\nu\in\PPP_T^{\lambda}$ the process
$\left(\log\left(Z^{\lambda-\delta,
\nu}_t\right)-A_t\right)_{t\in[0,T]}$ is a
$\QQ^{\lambda-\delta,\nu}$-submartingale and that there exists a
process $\hat{\nu}\in\PPP_T^{\lambda}$ for which
$\left(\log\left(Z^{\lambda-\delta,
\hat{\nu}}_t\right)-A_t\right)_{t\in[0,T]}$ is a true
$\QQ^{\lambda-\delta,\hat{\nu}}$-martingale. Hence,
$$\underset{\QQ\in\MM^e_T}{\essinf}H_t(\QQ,T)=\underset{\nu\in\PPP^{\lambda}_T}{\essinf}H_t(\QQ^{\lambda,\nu},T)=
h\left(\frac{1}{\gamma_t}\right)-\frac{A_t}{\gamma_t}$$ and the
infimum is attained by the measure $\QQ^{\lambda,\hat{\nu}}$. In
the view of \eqref{eq: proof1}, this means that we can restrict
the set of measures in $\PPP_T^H$, provided that
$\phi\in\PPP_T^H$. For the latter, we have that item (i) and an
application of H\"{o}lder's inequality guarantees the existence of
constant $p>1$ such that
$\EE\left[\exp\left(\frac{p}{2}\int_0^T(\phi^2_u+\lambda^2_u)du\right)\right]<\infty$
which implies not only that $\phi\in\PPP_T^{\lambda}$, but also
that $\QQ^{\lambda,\phi}$ has finite entropy with respect to $\PP$
(see among others Remark 1.2 in \cite{Kaz94}). Therefore,
$$H^{(0)}_t(T)=\underset{\nu\in\PPP^{H}_T}{\essinf}H_t(\QQ^{\lambda,\nu},T)=h\left(\frac{1}{\gamma_t}\right)-\frac{A_t}{\gamma_t}.$$
Now, from \eqref{eq:dual representation} we have that for every
martingale measure for every $\nu\in\PPP_T^{\lambda}$ and for all
$t\in[0,T]$
$$v^{(b)}_t(C)\leq\EE_{\QQ^{\lambda,\nu}}[C|\FF_t]+H_t(\QQ^{\lambda,\nu},T)-H^{(0)}_t(T)$$
Proposition \ref{pro: BSDE represantation} guarantees that there
is a triple $(v^{(b)}_t(C),\zeta_t,\theta_t)_{t\in[0,T]}$ that
solves the BSDE \eqref{eq: BSDE of price}. Thus, for every
 $\nu\in\PPP_T^{\lambda}$
\begin{equation}\label{eq: proof2}
  v^{(b)}_t(C) = \EE_{\QQ^{\lambda,\nu}}\left[\left.C+\frac{1}{2}\int_t^T(2\zeta_u(\nu_u-\phi_u)-\gamma_u\zeta^2_u)du
  -\int_t^T\zeta_udW^{2,\nu}_u-\int_t^T\delta_udW^{1,\lambda}_u\right\vert\FF_t\right]
\end{equation}
Note that for process $\nu^*$, it holds that
$2\zeta(\nu^*-\phi)-\gamma\zeta^2=\frac{(\nu^*-\phi)^2}{\gamma}.$
From Lemma \ref{lem:bounds} it follows that
$\left(\int_0^t\zeta_udW_u^{2,\phi}\right)_{t\in[0,T]}$ is a
$BMO(\QQ^{\lambda,\phi})$-martingale and by Theorem 3.1 of
\cite{Kaz94} and the H\"{o}lder's inequality we get that
$\nu^*\in\PPP_T^{\lambda}$. It is left to show that this minimizer
belongs in $\PPP_T^{H}$ too. For this, we first show that
$\left(\int_0^t\zeta_udW^{2,\nu^*}\right)_{t\in[0,T]}$ and
$\left(\int_0^t\delta_udW^{1,\lambda}\right)_{t\in[0,T]}$ are
$BMO(\QQ^{\lambda,\nu^*})$-martingales. Following the lines of the
proof of Lemma \ref{lem:bounds}, we apply It\^{o}'s formula for
the process $\left(e^{\beta C_t}\right)_{t\in[0,T]}$ (where
$\beta\in\R_+$ and $(C_t)_{t\in[0,T]}$ is the solution of
\eqref{eq: BSDE of price}) to get that for any stopping time
$\tau\in[0,T]$
$$e^{\beta C_T}-e^{\beta C_{\tau}}=\int_{\tau}^Te^{\beta C_t}\left(\frac{\zeta^2_t}{2}\beta(\beta-\gamma_t)
+\frac{\beta^2}{2}\delta^2_t\right)dt+ \int_{\tau}^Te^{\beta
C_t}\beta\zeta_tdW^{2,\nu^*}_t+\int_{\tau}^Te^{\beta
C_t}\beta\delta_tdW^{1,\lambda}_t.$$ Note that there is a sequence
of stopping times $\tau_n\nearrow T$ such that
$\int_{0}^{t\wedge\tau_n}\beta e^{\beta C_s}\zeta_sdW^{2,\nu^*}_s$
and $\int_{0}^{t\wedge\tau_n}\beta e^{\beta
C_s}\delta_sdW^{1,\lambda}_s$ are true
$\QQ^{\lambda,\nu^*}$-martingales. Boundedness of the claim $C$
implies that for every $n\in\N$
\begin{equation*}
    2e^{\beta ||C||_{\linf}}\geq \EE_{\QQ^{\lambda,\nu^*}}\left[\left.\int_{t\wedge\tau_n}^Te^{\beta C_t}\frac{\zeta^2_t}{2}\beta(\beta-\gamma_t)dt\right\vert\FF_{\tau\wedge\tau_n}\right]
+\EE_{\QQ^{\lambda,\nu^*}}\left[\left.\int_{t\wedge\tau_n}^Te^{\beta
C_t}\frac{\beta^2}{2}\delta_t^2dt\right\vert\FF_{\tau\wedge\tau_n}\right]
\end{equation*}
Hence for $\beta>K_{\gamma}$,
\begin{equation*}
    2e^{\beta ||C||_{\linf}}\geq e^{-\beta ||C||_{\linf}}\frac{\beta(\beta-K_{\gamma})}{2}\EE_{\QQ^{\lambda,\nu^*}}\left[\left.\int_{t\wedge\tau_n}^T\zeta^2_tdt\right\vert\FF_{\tau\wedge\tau_n}\right]
+e^{-\beta||C||_{\linf}}\frac{\beta^2}{2}\EE_{\QQ^{\lambda,\nu^*}}\left[\left.\int_{t\wedge\tau_n}^T\delta_t^2dt\right\vert\FF_{\tau\wedge\tau_n}\right]
\end{equation*}
Letting $n\longrightarrow+\infty$ and taking supremum with respect
to stopping times implies that
$\left(\int_0^t\zeta_udW^{2,\nu^*}\right)_{t\in[0,T]}$ and
$\left(\int_0^t\delta_udW^{1,\lambda}\right)_{t\in[0,T]}$ are
$BMO(\QQ^{\lambda,\nu^*})$-martingales.

For the finite entropy of measure $\QQ^{\lambda,\nu^*}$, it
remains to show that
$\EE_{\QQ^{\lambda,\nu^*}}\left[\int_0^T\phi_t^2dt\right]$ and
$\EE_{\QQ^{\lambda,\nu^*}}\left[\int_0^T\lambda_t^2dt\right]$ are
finite. Note that
\begin{equation}\label{eq:proof2a}
\EE_{\QQ^{\lambda,\nu^*}}\left[\int_0^T\phi_t^2dt\right]=
\EE_{\QQ^{\lambda,\phi}}\left[e^{-\frac{1}{2}\int_0^T\gamma_t^2\zeta_t^2dt-\int_0^T\gamma_t\zeta_tdW^{2,\phi}_t}\int_0^T\phi_t^2dt\right]
\end{equation}
Thanks to the Lemma \ref{lem:bounds} and the uniform boundedness
of $(\gamma_t)_{t\in[0,T]}$,
$\left(\int_0^t\gamma_u\zeta_udW^{2,\phi}\right)_{t\in[0,T]}$ is
$BMO(\QQ^{\lambda,\phi})$-martingale. By Theorem 3.1 of
\cite{Kaz94}, there exists a constant $p>1$ such that
\begin{equation}\label{eq: proof 3}
    \EE_{\QQ^{\lambda,\phi}}\left[e^{-\frac{p}{2}\int_0^T\gamma_t^2\zeta_t^2dt-p\int_0^T\gamma_t\zeta_tdW^{2,\phi}_t}\right]<\infty
\end{equation}
If we apply H\"{o}lder's inequality in \eqref{eq:proof2a}, we get
that its terms are finite if and only if
$\EE_{\QQ^{\lambda,\phi}}\left[\left(\int_0^T\phi_t^2dt\right)^q\right]$
is finite, where $\frac{1}{p}+\frac{1}{q}=1$. However, this is
guaranteed by Assumption \ref{ass:1} and item (i).

The same arguments proves that
$\EE_{\QQ^{\lambda,\nu^*}}\left[\int_0^T\lambda_t^2dt\right]<\infty$.
\end{proof}

\begin{remark}
The martingale measure which minimizes the penalty process
$\alpha_{t,T}$, is the measure $\QQ^{\lambda,\phi}$, which does
not depend on the time horizon $T$. This means that the agent's
marginal utility valuation (the so-called Davis price) is the
(conditional) expectation of the payoff under the same martingale
measure regardless the maturity of the claim. This is in contrast
with the backward exponential valuation, where the corresponding
measure is the minimal entropy martingale measure, which depends
on the time horizon the utility lives in.
\end{remark}

We can now take advantage of representation \eqref{eq:robust
repre} and prove some properties of the valuation under forward
exponential performance criteria, where we use the notation
$v^{(b)}_t(C;\gamma)$ to emphasize (when needed) the dependence of
the indifference valuation on the risk aversion process.

\begin{proposition}\label{pro:price properties}
Impose the conditions of Theorem \ref{thm:robust repre} and fix a
time horizon $T>0$. For every contingent claim $C\in\linf(\FF_T)$
the following statements hold true.
\begin{itemize}
    \item [(i)] The forward (buyer) indifference value is decreasing with
    respect to risk aversion in the following sense:
If $(g_t)_{t\in[0,T]}$ is another risk aversion process with
$\gamma_T\geq g_T$, a.s., then $$v^{(b)}_t(C;\gamma)\leq
v^{(b)}_t(C;g)\,\,\,\,\text{a.s.},$$ for any $t\in [0,T]$.
    \item [(ii)] Let $\left((\gamma_t(n))_{t\in[0,T]}\right)_{n\in\N}$ be a sequence of risk aversion processes, such that $\gamma_T(n)\nearrow\infty$ in $\PP$,
    as $n\rightarrow\infty$. Then $$v^{(b)}_{t}(C;
\gamma(n))\longrightarrow\underset{\nu\in\PPP_T^{H}}{\inf}\EE_{\QQ^{\lambda,\nu}}[C|\FF_t]\,\,\,\,\text{a.s.},$$
for any $t\in [0,T]$.

Also if $\gamma_T(n)\searrow 0$ in $\PP$,
    as $n\rightarrow\infty$ then, $$v^{(b)}_{t}(C;
\gamma(n))\longrightarrow\EE_{\QQ^{\lambda,\phi}}[C|\FF_t]\,\,\,\,\text{a.s.},$$
for any $t\in [0,T]$.
    \item [(iii)] For each risk aversion coefficient, the forward indifference valuation is time consistent in the sense that
    $$v^{(b)}_{\tau}(v^{(b)}_s(C))=v^{(b)}_{\tau}(C)\,\,\,\,\text{a.s.},$$
for any stopping times $\tau,s$ with $\tau\leq s$.
    \item[(iv)] The indifference valuation is replication invariance, i.e. $$v^{(b)}_t(C+X_T^{\theta})=v^{(b)}_t(C)+X_t^{\theta}\,\,\,\,\text{a.s.},$$
    for every $\theta\in\AA$.
\end{itemize}
\end{proposition}
\begin{proof}

Part (i) follows directly from the robust representation of the
forward indifference valuation given in \eqref{eq:robust repre}.
Again from \eqref{eq:robust repre} and the monotone convergence
theorem we get the first item of part (ii). For the limit of the
indifference value when $\gamma_T(n)\searrow 0$, we use similar
arguments as the ones in \cite{ManSch05}. Thanks to the robust
representation of the $v_t^{(b)}(C;\gamma(n))$, it is enough to
show that
$$\underset{n\rightarrow\infty}{\liminf}v_t^{(b)}(C;\gamma(n))\geq\EE_{\QQ^{\lambda,\phi}}[C|\FF_t].$$
We shall show the above inequality for $t=0$, since the more
general case is proved similarly. By Fenchel-Young inequality
$xp\geq\frac{p-e^{-ax}}{a}-\frac{p\log(p)}{a}$, we get that for
every $\nu\in\PPP^{H}_T$
\begin{eqnarray*}
  \EE_{\QQ^{\lambda,\nu}}[C]=\EE_{\QQ^{\lambda,\phi}}\left[C\frac{Z^{0,\nu}_T}{Z^{0,\phi}_T}\right] &\geq &
  \EE_{\QQ^{\lambda,\phi}}\left[\frac{\frac{Z^{0,\nu}_T}{Z^{0,\phi}_T}-e^{-\gamma_T(n)C}}{\gamma_T(n)}
  -\frac{1}{\gamma_T(n)}\frac{Z^{0,\nu}_T}{Z^{0,\phi}_T}\log\left(\frac{Z^{0,\nu}_T}{Z^{0,\phi}_T}\right)\right]
    \\
  &=& \EE_{\QQ^{\lambda,\phi}}\left[\frac{\frac{Z^{0,\nu}_T}{Z^{0,\phi}_T}-e^{-\gamma_T(n)C}}{\gamma_T(n)}\right]-
  \EE_{\QQ^{\lambda,\nu}}\left[\frac{1}{\gamma_T(n)}\frac{Z^{0,\nu}_T}{Z^{0,\phi}_T}\log\left(\frac{Z^{0,\nu}_T}{Z^{0,\phi}_T}\right)\right]\\
  &=& \frac{1}{\gamma_0(n)}-\EE_{\QQ^{\lambda,\phi}}\left[\frac{e^{-\gamma_T(n)C}}{\gamma_T(n)}\right]-
  \frac{1}{2}\EE_{\QQ^{\lambda,\nu}}\left[\frac{1}{\gamma_T(n)}\int_0^T(\phi_s-\nu_s)^2ds\right]\\
  &=& \EE_{\QQ^{\lambda,\phi}}\left[\frac{1-e^{-\gamma_T(n)C}}{\gamma_T(n)}\right]-
  \frac{1}{2}\EE_{\QQ^{\lambda,\nu}}\left[\frac{1}{\gamma_T(n)}\int_0^T(\phi_s-\nu_s)^2ds\right]
\end{eqnarray*}
Thus,
$$\underset{n\rightarrow\infty}{\liminf}v_0^{(b)}(C;\gamma(n))\geq
\underset{n\rightarrow\infty}{\liminf}\EE_{\QQ^{\lambda,\phi}}\left[\frac{1-e^{-\gamma_T(n)C}}{\gamma_T(n)}\right],$$
which gives the desired result.

Part (iii) follows from the definition of the forward indifference
valuation \eqref{eq:indifference price} and a simple application
of the dynamic programming principle. Finally, part (iv) is a
consequence of the indifference valuation robust representation
\eqref{eq:robust repre}.
\end{proof}

\begin{remark}\label{rem:back vs forward}
All items of Proposition \ref{pro:price properties} can be
considered as extensions of the properties of the backward
indifference value $pr_t^{(b)}(C)$, defined through exponential
utility function in \eqref{eq:back indifference price}. For
example, although the forward performance criterion has stochastic
risk aversion, the monotonicity of the indifference value is
preserved, something that is consistent with the financial
intuition: the higher the risk aversion at the time of maturity
is, the lower price the buyer is going to bid.

The main difference between forward and backward valuation is that
the so-called marginal martingale probability measure, i.e.~the
measure that minimizes penalty function, is $\QQ^{\lambda,\phi}$
in forward valuation and the minimal entropy martingale measure in
the backward case. The important difference between these measures
is that $\QQ^{\lambda,\phi}$ does not depend on a time horizon.
\end{remark}

\begin{remark}\label{rem:constant gamma}
A special case of the forward exponential performance is when the
risk aversion is constant. Then, the problem of the indifference
valuation has an immediate relation with the associated problem of
the backward valuation. This is because, given a maturity $T$ of a
contingent claim, the forward utility function at $T$ can be
written as $U_T(x)=-e^{-\gamma(x-\frac{A_T}{\gamma})}$, and the
term $-\frac{A_T}{\gamma}$ can be thought as an $\FF_T$-measurable
random endowment. Therefore, we may consider the forward
performance indifference valuation as a backward indifference
valuation under this random endowment. A number of properties of
this value, called conditional indifference price, have been
provided in the Appendix of \cite{AntZit10}.
\end{remark}

One further property of the indifference value that can be proved
using the robust representation of the price is the following.

\begin{proposition}\label{pro:derivative}
Impose the conditions of Theorem \ref{thm:robust repre} and let
$C\in\linf(\FF_T)$ be a contingent claim for some time horizon
$T>0$. Then, the function $$\R\ni a\mapsto f(a):=v_0^{(b)}(aC)$$
is differentiable and
\begin{equation}\label{eq:derivative}
    f'(a)=\EE_{\QQ^{\lambda,\phi^{(aC)}}}[C]
\end{equation}
where $\phi_t^{(aC)}=\phi_t-\gamma_t\zeta_t(aC)$, for $t\in[0,T]$
and $\zeta(aC)$ is the corresponding part of the solution of
\eqref{eq: BSDE of price} for boundary condition $aC$.
\end{proposition}
\begin{proof}
We will show the result when $a=0$. Thanks to item (ii) of
Proposition \ref{pro:price properties}, we have
\begin{eqnarray*}
  \underset{\epsilon\searrow 0}{\lim}\frac{v_0^{(b)}(\epsilon C)}{\epsilon} &=&
  \underset{\epsilon\searrow 0}{\lim}\underset{\nu\in\PPP_T^{H}}{\min}\EE_{\QQ^{\lambda,\nu}}\left[C+\frac{1}{2\epsilon\gamma_T}\int_0^T(\nu_u-\phi_u)^2du\right] \\
   &=& \EE_{\QQ^{\lambda,\phi}}\left[C\right]
\end{eqnarray*}
We also observe that
$\underset{\epsilon\nearrow0}{\lim}\frac{v_0^{(b)}(\epsilon
C)}{\epsilon}=-\underset{\epsilon\searrow
0}{\lim}\frac{v_0^{(b)}(-\epsilon
C)}{\epsilon}=\EE_{\QQ^{\lambda,\phi}}\left[C\right]$, which means
that $f'(0)=\EE_{\QQ^{\lambda,\phi}}\left[C\right]$.

The more general case of $a\neq 0$ follows from the same arguments
and Corollary \ref{cor:new phi}.
\end{proof}
Note that in the case of backward valuation the situation is
similar. Namely, the function $g(a)=pr_0^{(b)}(aC)$ is also
differentiable and its derivative is given as $\EE_{\QQ(aC)}[C]$,
where the $\QQ(aC)$ is the martingale measure that minimizes the
relative entropy with risk to the measure $\PP(aC)$ defined by its
R-N derivative $\frac{d\PP(aC)}{d\PP}=ce^{-\gamma aC}$, for the
appropriate constant $c$ (see \cite{IlJonSir05} for details on
this result).
\begin{remark}
The derivative of the indifference valuation with respect to the
units of a given claim can be used in the determination of the
number of units that the agent is willing to sell/buy when the
price of the contingent claim is given. In other words, it leads
to the agent's demand function on this claim in the same manner as
in \cite{AntZit10}. This differentiation result can be applied for
a vector of claims in straightforward way.
\end{remark}

The arguments of the proposition below follow similar lines as
those in Proposition 14 of \cite{ManSch05}.

\begin{proposition}\label{pro:convergence of the price}
Impose the conditions of Theorem \ref{thm:robust repre} and let
$C^n$ be a bounded sequence in $\linf(\FF_T)$ such that
$C^n\rightarrow C$ in probability for some $C\in\linf(\FF_T)$.
Then
\begin{equation}\label{eq:convergence}
    \underset{0\leq t\leq
    T}{\sup}|v^{(b)}_t(C^n)-v^{(b)}_t(C)|\rightarrow 0
\end{equation}
in probability.
\end{proposition}
\begin{proof}
Thanks to Proposition \ref{pro: BSDE represantation}, the
indifference values of $C$ and $C^n$ satisfy the following
relations
\begin{eqnarray*}
  C &=& v_t^{(b)}(C)+\frac{1}{2}\int_t^T\gamma_u\zeta^2_udu+\int_t^T\zeta_udW^{2,\phi}_u+\int_t^T\theta_u\sigma(Y_u)dW^{1,\lambda}_u \\
  C^n &=& v_t^{(b)}(C^n)+\frac{1}{2}\int_t^T\gamma_u(\zeta_u^n)^2du+\int_t^T\zeta_u^ndW^{2,\phi}_u+\int_t^T\theta_u^n\sigma(Y_u)dW^{1,\lambda}_u
\end{eqnarray*}
for some processes $\zeta,\zeta^n,\theta$ and $\theta^n$. Hence,
\begin{equation}\label{eq:limit 1}
v_t^{(b)}(C^n)-v_t^{(b)}(C)=C^n-C+\int_t^T(\theta_u-\theta_u^n)\sigma(Y_u)dW^{1,\lambda}_u+\frac{1}{2}\int_t^T\gamma_u(\zeta_u^2-(\zeta_u^n)^2)du+\int_t^T(\zeta_u-\zeta_t^n)dW^{2,\phi}_u.
\end{equation}
We then define for each
$n\in\N$, the sequence of processes
$\nu_t(n)=-\frac{1}{2}\gamma_t(\zeta_t+\zeta_t^n)$ and by Lemma
\ref{lem:bounds} $\nu(n)\in\PPP^{\lambda}_T$ for each $n$. Under
the probability measure $\QQ^{(n)}$ defined through its R-N
derivative
$$\frac{d\QQ^{(n)}}{d\QQ^{\lambda,\phi}}=\mathcal{E}_T\left(\int_0^{.}\nu_u(n)dW^{2,\phi}_u\right)$$
we have that
\begin{equation}\label{eq:limit 2}
v_t^{(b)}(C^n)-v_t^{(b)}(C)=\EE_{\QQ^{(n)}}[C^n-C|\FF_t].
\end{equation}
The next step is to show that the process
$\left(\int_0^t\gamma_u(\zeta_u+\zeta_u(n))dW^{2,\phi}_u\right)_{t\in[0,T]}$
is a $BMO(\QQ^{\lambda,\phi})$-martingale. Indeed, thanks to the
Lemma \ref{lem:bounds} we have that
  $$\left\vert\left\vert \int_0^\cdot\gamma_u(\zeta_u+\zeta_u(n))dW^{2,\phi}_u\right\vert\right\vert_{BMO}^2 =
  \underset{\tau}{\sup}\left\vert\left\vert
  \EE_{\QQ^{\lambda,\phi}}\left[\left.\int_{\tau}^T\gamma_u^2(\zeta_u+\zeta_u(n))^2du\right\vert\FF_{\tau}\right]\right\vert\right\vert_{\linf}$$
    $$\leq K_{\gamma}^2\underset{\tau}{\sup}\left\vert\left\vert
   \EE_{\QQ^{\lambda,\phi}}\left[\left.\int_{\tau}^T(\zeta_u(n))^2du\right\vert\FF_{\tau}\right]\right\vert\right\vert_{\linf}+
   K_{\gamma}^2\underset{\tau}{\sup}\left\vert\left\vert
   \EE_{\QQ^{\lambda,\phi}}\left[\left.\int_{\tau}^T(\zeta_u)^2du\right\vert\FF_{\tau}\right]\right\vert\right\vert_{\linf}<\infty$$
where supremum is taken under all stopping times $\tau$ in
$[0,T]$. Theorem 3.1 of \cite{Kaz94} guarantees the existence of
two constants $p>1$ and $c>0$ such that
$$\underset{0\leq t\leq T}{\sup}\EE_{\QQ^{\lambda,\phi}}\left[\left.e^{-\frac{p}{2}\int_{t}^T\nu(n)_u^2du-p\int_{t}^T\nu(n)_udW^{2,\phi}_u}\right\vert\FF_{t}\right]<c$$
for each $n\in\N$. Then, \eqref{eq:convergence} follows by
applying the H\"{o}lder's and the Doob's maximal inequalities.
\end{proof}

\bigskip

\section{Optimal Risk Sharing}

In the present section, we consider two financial agents whose
investment criteria are modelled by forward exponential
performances and we address the problem of \textit{optimal risk
sharing}. We denote the characterization pairs of agents'
performance criteria by $(\tone_t,\phione_t)_{t\in[0,\infty)}$ and
$(\ttwo_t,\phitwo_t)_{t\in[0,\infty)}$, with the risk aversion
processes $\gone$, $\gtwo$ defined by the equations
$\frac{1}{\gone}=X^{\frac{1}{\gone_0},\tone}$ and
$\frac{1}{\gtwo}=X^{\frac{1}{\gtwo_0},\ttwo}$. Also, $U^{i}_t$
stands for the corresponding agent's forward utility at time $t$
and $\Ai$ is the associated characterization process, for $i=1,2$.
In addition, we assume that each agent has some initial
(non-replicable) endowment in her portfolio, denoted by
$\EN^i\in\linf(\FF_T)$, for some time horizon $T>0$. The sum
$\EN=\EN^1+\EN^2$ is the so-called \textit{aggregated random
endowment}.

The problem of optimal risk sharing (as formed in the mathematical
finance literature in \cite{BarElk05}) is finding a contract $C^*$
and a price $p^*$ which solve the following problem
\begin{eqnarray*}
  & & \underset{C,p}{\argmax}\,\underset{\pi\in\AA^{\infty}}{\sup}\EE\left[U^{1}_T(\EN^1+\int_0^T\pi_sdS_u+C-p)\right]   \\
  & & \text{Given that}\\
  & &
  \underset{\pi\in\AA^{\infty}}{\sup}\EE\left[U^{2}_T(\EN^2+\int_0^T\pi_sdS_u)\right]\leq
  \underset{\pi\in\AA^{\infty}}{\sup}\EE\left[U^{2}_T(\EN^2+\int_0^T\pi_sdS_u-C+p)\right]
\end{eqnarray*}

From the definition of the indifference valuation
\eqref{eq:indifference price} and its replication invariance
property (part (iv) of Proposition \ref{pro:price properties}) we
get the following more tractable equivalent problem
\begin{equation}\label{eq:risk sharing problem}
    \underset{C}{\argmax}\{v^{1}_0(\EN^1+C)+v^{2}_0(\EN^2-C)\}
\end{equation}
where, $v^{i}_0(\cdot)$ denotes the (buyer) indifference valuation
of the corresponding agent at time $t=0$, where $i=1,2$.
\begin{definition}\label{def:pareto}
We say that agents are in Pareto optimal situation if the set of
the solutions of problem \eqref{eq:risk sharing problem} is the
set of replicable claims or equivalently if
\begin{equation}\label{eq:pareto}
v_0^{1}(\EN^1+C)+v_0^{2}(\EN^2-C)< v_0^{1}(\EN^1)+v_0^{2}(\EN^2).
\end{equation}
for every $C$ which is not replicable.
\end{definition}
The inequality \eqref{eq:pareto} introduces the so-called
\textit{inf-convolution measure}
\begin{equation}\label{eq:inf-convolution}
    \rho(\EN)=\underset{C}{\inf}\{\rho^{1}(\EN-C)+\rho^{2}(C)\}
\end{equation}
where $\rho^{i}(C)=-v_0^{i}(C)$ for $i=1,2$ is the convex risk
measure induced by the associated forward exponential performance
criteria. Note that $\rho(\cdot)$ maps payoffs in $\linf(\FF_T)$
to $\R\cup\{-\infty\}$, where the time horizon $T$ is the maturity
of the agents' endowments.

\begin{assumption}\label{ass:2}
There exist constants $\epsilon, K>0$ such that
$\underset{t\in[0,T]}{\sup}||\gone_t||_{\linf},\underset{t\in[0,T]}{\sup}||\gtwo_t||_{\linf}<K$
and $\EE[e^{(1+\epsilon)\int_0^T\phione^2_udu}],
\EE[e^{(1+\epsilon)\int_0^T\phitwo^2_udu}]<\infty$.
\end{assumption}

A first result, the proof of which is based on Theorem
\ref{thm:robust repre}, is that the inf-convolution measure of two
forward exponential performance processes is \textit{not} a risk
measure that is induced by another forward exponential performance
criterion (something which is in contrast with the inf-convolution
measure induced by exponential utility functions).

\begin{proposition}\label{pro:inf-convo}
Impose Assumptions \ref{ass:1} and \ref{ass:2}. The
inf-convolution risk measure $\rho(\cdot)$ defined in
\eqref{eq:inf-convolution} is induced by a forward exponential
performance process if and only if $\phione_t=\phitwo_t$ for all
$t\in[0,T]$, a.s. In this case, the characterization pair of the
forward exponential performance is
$(\tone_t+\ttwo_t,\phione_t)_{t\in[0,T]}$, i.e.~the risk aversion
process is given by
$$\Gamma_t=\frac{\gone_t\gtwo_t}{\gone_t+\gtwo_t}$$
and the characterization process by
$$A_t=\frac{1}{2}\int^t_0\left(\lambda(Y_u)-\delta_u\right)^2du
-\frac{1}{2}\int^t_0\phi^2_udu-
    \int^t_0\phi_udW^2_u.$$
where $\delta_t=\Gamma_t(\tone_t+\ttwo_t)\sigma(Y_t)$.
\end{proposition}

\begin{proof}
Theorem 3.6 in \cite{BarElk05} states that the penalty function of
the inf-convolution measure is the sum of the penalty function of
the involved risk measures. In our setting, the penalty function
of $\rho(\cdot)$ can be written as
\begin{equation}\label{eq:help1}
\alpha_{t,T}(\QQ^{\lambda,\nu})=\frac{1}{2}\EE_{\QQ^{\lambda,\nu}}\left[\left.
\int_t^T\frac{(\nu_s-\phione_s)^2}{\gone_s}+\frac{(\nu_s-\phitwo_s)^2}{\gtwo_s}ds\right\vert\FF_t\right]
\end{equation}
for every $t\in[0,T]$.

Assume that $\rho(\cdot)$ is induced by a forward exponential
performance. A necessary condition for this is the existence of a
risk aversion process $\Gamma$ and a process $\tilde{\phi}$ such
that
$$\EE_{\QQ^{\lambda,\nu}}\left[\left.
\int_t^T\frac{(\nu_s-\tilde{\phi}_s)^2}{\Gamma_s}ds\right\vert\FF_t\right]=\EE_{\QQ^{\lambda,\nu}}\left[\left.
\int_t^T\frac{(\nu_s-\phione_s)^2}{\gone_s}+\frac{(\nu_s-\phitwo_s)^2}{\gtwo_s}ds\right\vert\FF_t\right]$$
for every $t\in[0,T]$ and for every process $\nu\in\PPP_T^{H}$.
Setting $\nu=\tilde{\phi}$ and taking into account the positivity
of $\gone$ and $\gtwo$, we get that $\phione_t=\phitwo_t$ for all
$t\in[0,T]$, a.s. But
$\left(\frac{\gone+\gtwo}{\gone\gtwo}\right)$ is replicable and
bounded, therefore
\begin{equation*}
\EE_{\QQ^{\lambda,\nu}}\left[\left.
\frac{1}{\Gamma_T}\int_t^T(\nu_s-\tilde{\phi}_s)^2ds\right\vert\FF_t\right]=\EE_{\QQ^{\lambda,\nu}}\left[\left.
\frac{\gone_T+\gtwo_T}{\gone_T\gtwo_T}\int_t^T(\nu_s-\phi_s)^2ds\right\vert\FF_t\right]
\end{equation*}
for every $t\in[0,T]$ and for every process $\nu\in\PPP_T^{H}$,
which first implies that $\tilde{\phi}=\phione$ and then
$\Gamma_t=\frac{\gone_t\gtwo_t}{\gone_t+\gtwo_t}$, $\forall
t\in[0,T]$.

For the inverse part, we assume that $\phione_t=\phitwo_t$ for all
$t\in[0,T]$, a.s. Equation \eqref{eq:help1} implies that
$$\alpha_{t,T}(\QQ^{\lambda,\nu})=\frac{1}{2}\EE_{\QQ^{\lambda,\nu}}\left[\left.
\int_t^T\frac{\gone_s+\gtwo_s}{\gone_s\gtwo_s}(\nu_s-\phi_s)^2ds\right\vert\FF_t\right]$$
for every $t$ and $\nu$. Letting
$\Gamma_t=\frac{\gone_t\gtwo_t}{\gone_t+\gtwo_t}$ completes the
proof.
\end{proof}

\begin{remark}
Proposition \ref{pro:inf-convo} states that only in the case where
agents adapt their subjective probability measure up to the
maturity of their endowments in the same manner, the
representative agent can be considered as behaving under forward
exponential performance criteria. In other word, when
$\phione=\phitwo$ Theorem 2.3 in \cite{BarElk05} has a direct
extension in the case of stochastic risk aversion.
\end{remark}

\subsection{The special case of replicable endowments}

A special case is when agents do not have any endowment in their
initial portfolios or when both endowments are replicable, that is
$\exists\pi^1,\pi^2\in\AA^{\infty}$ and $c^1,c^2\in\R$ such that
$\EN^i=X_T^{c^i,\pi^i}$, for $i=1,2$.

\begin{proposition}\label{pro:Pareto and measure}
Let Assumptions \ref{ass:1} and \ref{ass:2} hold true and assume
that $\EN^1$ and $\EN^2$ are replicable. Then, agents are in
Pareto optimal situation if and only if $\phione_t=\phitwo_t$ for
every $t\in[0,\infty)$ a.s.
\end{proposition}

\begin{proof}
The result follows from the dual representation of the
indifference valuation \eqref{eq:dual representation} and the
Theorem 3.6 of \cite{BarElk05}, which states that Pareto
optimality is equivalent to the equality of the agents' marginal
martingale measures, that is
$\QQ^{\lambda,\phione}=\QQ^{\lambda,\phitwo}$.
\end{proof}

Proposition \ref{pro:Pareto and measure} implies that agents are
willing to trade some non-replicable claims if and only if the way
they adapt their subjective probability measure is not the same at
all times. Note that this statement is independent on the agents'
risk aversion processes $\gone$ and $\gtwo$. Another way to
interpret this result is that if agents do not include in their
utilities the unhedgeable part of the market (i.e.~when
$\phione=\phitwo=0$), they are unwilling to make any
non-replicable transaction, no matter how their risk aversion
processes differ to each other. This can be seen as a
generalization of the corresponding result in the case of backward
exponential utility (see Proposition 3.8 in \cite{AntZit10}).

\subsection{The case of constant risk aversions}\label{subsec:constant gamma}

In the simplified case where the agents' risk aversion are
constant, we can explicitly solve the sharing problem
\eqref{eq:risk sharing problem}. This is in fact because under
constant risk aversion, the contingent claim valuation problem can
be written as conditional indifference valuation under classical
exponential utility (see Remark \ref{rem:constant gamma}).

\begin{proposition}\label{pro:optimal sharing}
Impose Assumptions \ref{ass:1} and \ref{ass:2} and suppose in
addition that agents' risk aversions are constants,
i.e.~$\gone,\gtwo\in\R_+$. Any claim of the form
\begin{equation}\label{eq:optimal claim1}
    \frac{1}{\gone+\gtwo}\left(\int_0^T\frac{\phitwo_t^2-\phione_t^2}{2}dt+
\int_0^T(\phitwo_t-\phione_t)dW^2_t
\right)+\frac{\gtwo\EN^2-\gone\EN^1}{\gone+\gtwo}+X_T^{c,\pi}
\end{equation}
for some constant $c$ and admissible strategy $\pi\in\AA$, solves
the optimal sharing problem \eqref{eq:risk sharing problem}.
\end{proposition}

\begin{proof}
By robust representation \eqref{eq:robust repre} (see also its
proof), we get that for every claim $C$ it holds that
\begin{equation}\label{ineq: prices}
v^{1}_0(\EN^1+C)+v^{2}_0(\EN^2-C)\leq
\underset{\nu\in\PPP_T^{H}}{\inf}\EE_{\QQ^{\lambda,\nu}}\left[\EN
+\frac{1}{2\gone}\int_0^T(\nu_s-\phione_s)^2ds+\frac{1}{2\gtwo}\int_0^T(\nu_s-\phitwo_s)^2ds\right]
\end{equation}
For very claim $C^*$ of the form \eqref{eq:optimal claim1}, we
have that
\begin{eqnarray*}
 v^{1}_0(\EN^1+C^*) &=& \underset{\nu\in\PPP_T^{H}}{\inf}\EE_{\QQ^{\lambda,\nu}}\left[\frac{\gtwo}{\gone+\gtwo}\EN
+\frac{1}{2}\int_0^T\left(\frac{(\nu_s-\phione_s)^2}{\gone}+\frac{\phitwo_s^2-\phione_s^2}{\gone+\gtwo}-\frac{2(\phitwo_s-\phione_s)}{\gone+\gtwo}\right)ds\right]+c \\
   &=& \frac{\gtwo}{\gone+\gtwo}\underset{\nu\in\PPP_T^{H}}{\inf}\EE_{\QQ^{\lambda,\nu}}\left[\EN
+\frac{1}{2\gone}\int_0^T(\nu_s-\phione_s)^2ds+\frac{1}{2\gtwo}\int_0^T(\nu_s-\phitwo_s)^2ds\right]+c
\end{eqnarray*}
Similarly, we get that
\begin{eqnarray*}
 v^{2}_0(\EN^2-C^*) &=& \underset{\nu\in\PPP_T^{H}}{\inf}\EE_{\QQ^{\lambda,\nu}}\left[\frac{\gone}{\gone+\gtwo}\EN
+\frac{1}{2}\int_0^T\left(\frac{(\nu_s-\phitwo_s)^2}{\gtwo}-\frac{\phitwo_s^2-\phione_s^2}{\gone+\gtwo}+\frac{2(\phitwo_s-\phione_s)}{\gone+\gtwo}\right)ds\right]-c \\
   &=& \frac{\gone}{\gone+\gtwo}\underset{\nu\in\PPP_T^{H}}{\inf}\EE_{\QQ^{\lambda,\nu}}\left[\EN
+\frac{1}{2\gone}\int_0^T(\nu_s-\phione_s)^2ds+\frac{1}{2\gtwo}\int_0^T(\nu_s-\phitwo_s)^2ds\right]-c
\end{eqnarray*}
Therefore,
$$v^{1}_0(\EN^1+C^*)+v^{2}_0(\EN^2-C^*)=\underset{\nu\in\PPP_T^{H}}{\inf}\EE_{\QQ^{\lambda,\nu}}\left[\EN
+\frac{1}{2\gone}\int_0^T(\nu_s-\phione_s)^2ds+\frac{1}{2\gtwo}\int_0^T(\nu_s-\phitwo_s)^2ds\right]$$
which together with \eqref{ineq: prices} completes the proof.
\end{proof}
 In other words, the optimal risk sharing part consists of three
elements: the optimal sharing of the agents' random endowments
which is exactly the same as in the backward exponential utility
case (see \cite{BarElk04}), the sharing of the agents'
perspectives about the probability measure (in the way they are
incorporated on the agents' forward performances) and a replicable
part (which can essentially be ignored since it does not transfer
any risk).

If there are no endowments, the agents share their differences of
beliefs regarding the evolution of the probability measure through
the contract with payoff
$\frac{1}{\gone+\gtwo}\left(\int_0^T\frac{\phitwo_t^2-\phione_t^2}{2}dt+
\int_0^T(\phitwo_t-\phione_t)dW^2_t \right)$. Note that the
expectation of this payoff increases (in absolute terms) as the
differences of the processes $\phione$ and $\phitwo$ increases
(this means that intense difference in beliefs implies high volume
of transaction).

\subsection{The case of stochastic risk aversions}

In the case where the agents' risk aversion coefficients are
stochastic, the optimal sharing problem is more involved, since
the methods used in the backward exponential utility case can not
be applied. Recall that problem \eqref{eq:risk sharing problem} is
equivalent to finding a claim $C^*$ that maximizes the sum
$v^{1}_0(\EN^1+C)+v^{2}_0(\EN^2-C)$, for which we have
\begin{eqnarray*}
  & & v^{1}_0(\EN^1+C)+v^{2}_0(\EN^2-C) =   \\
  &=& \underset{\nu\in\PPP_T^{H}}{\inf}\EE_{\QQ^{\lambda,\nu}}\left[\EN^1+C +\frac{1}{2}\int_0^T\frac{(\nu_s-\phione_s)^2}{\gone_s}ds\right]
 + \underset{\nu\in\PPP_T^{H}}{\inf}\EE_{\QQ^{\lambda,\nu}}\left[\EN^2-C
 +\frac{1}{2}\int_0^T\frac{(\nu_s-\phitwo_s)^2}{\gtwo_s}ds\right]\\
\end{eqnarray*}
$$=\underset{\nu\in\PPP_T^{H}}{\inf}\EE_{\QQ^{\lambda,\nu}}\left[\EN+C
+\frac{1}{2}\int_0^T\frac{(\nu_s-\phione_s)^2}{\gone_s}ds\right]
 +
\underset{\nu\in\PPP_T^{H}}{\inf}\EE_{\QQ^{\lambda,\nu}}\left[-C
 +\frac{1}{2}\int_0^T\frac{(\nu_s-\phitwo_s)^2}{\gtwo_s}ds\right]$$

In the special case where the agents have the same stochastic risk
aversion process, the optimal risk sharing problem \eqref{eq:risk
sharing problem} can be solved explicitly.

\begin{proposition}\label{pro:same risk aversion}
Impose Assumptions \ref{ass:1} and \ref{ass:2} and suppose that
$\gone_t=\gtwo_t$ for every $t\in[0,T]$. Then, any claim of the
form
\begin{equation}\label{eq:optimal claim}
\frac{1}{2}\int_0^T\frac{\phitwo_t^2-\phione_t^2}{\gamma_t}dt+
\int_0^T\frac{\phitwo_t-\phione_t}{\gamma_t}dW^2_t+\frac{\EN^2-\EN^1}{2}+X_T^{c,\pi}
\end{equation}
for some constant $c$ and admissible strategy $\pi\in\AA$, solves
the optimal sharing problem \eqref{eq:risk sharing problem}.
\end{proposition}

\begin{proof}
It is enough to show that for every claim $C^*$ of the form
\eqref{eq:optimal claim}
\begin{equation}\label{eq:enough inequality}
v^{1}_0(\EN^1+C^*)+v^{2}_0(\EN^2-C^*)=
\underset{\nu\in\PPP_T^{H}}{\inf}\EE_{\QQ^{\lambda,\nu}}\left[
\EN+\frac{1}{2\gamma_T}\int_0^T\left((\nu_s-\phione_s)^2+(\nu_s-\phitwo_s)^2\right)ds\right]
\end{equation}
We first observe that for every such $C^*$ and every
$\nu\in\PPP_T^{H}$, $\EE_{\QQ^{\lambda,\nu}}\left[\EN^1+C^*
+\frac{1}{2}\int_0^T\frac{(\nu_s-\phione_s)^2}{\gamma_s}ds\right]$
equals to
\begin{eqnarray*}
    &  & \EE_{\QQ^{\lambda,\nu}}\left[\frac{\EN}{2}+\frac{1}{2\gamma_T}\left(\int_0^T((\nu_s-\phione_s)^2+\phitwo_s^2-\phione_s^2)ds+\int_0^T(\phitwo_s-\phione_s)dW_s^2\right)\right]\\
    & = & \EE_{\QQ^{\lambda,\nu}}\left[\frac{\EN}{2}+\frac{1}{4\gamma_T}\left(\int_0^T((\nu_s-\phione_s)^2+(\nu_s-\phitwo_s)^2)ds\right)\right]
\end{eqnarray*}
Applying the same calculations to
$\EE_{\QQ^{\lambda,\nu}}\left[\EN^2-C^*
+\frac{1}{2}\int_0^T\frac{(\nu_s-\phitwo_s)^2}{\gamma_s}ds\right]$,
we get that
$$v^{1}_0(\EN^1+C^*)+v^{2}_0(\EN^2-C^*) = 2\underset{\nu\in\PPP_T^{H}}{\inf}\EE_{\QQ^{\lambda,\nu}}\left[\frac{\EN}{2}+\frac{1}{4\gamma_T}\left(\int_0^T((\nu_s-\phione_s)^2+(\nu_s-\phitwo_s)^2)ds\right)\right]$$
which is equivalent to \eqref{eq:enough inequality}.
\end{proof}

If the agents have the same risk aversion process, the sharing
 of the endowments is exactly the same as the corresponding situation of
entropic risk measures (i.e., when agents have the same risk
aversion coefficient). This implies that after the transaction
both agents have the same random endowment, $\EN/2$. Note also the
similarity of the other terms of the optimal contract with those
of the case analyzed in subsection \ref{subsec:constant gamma}.

For the more general case of different and stochastic risk
aversion processes, the above arguments can not be applied.
However, we are able to construct the dynamics of the
inf-convolution measure, which gives the existence and an implicit
form of the optimal risk sharing contract.

We first note that under Assumptions \ref{ass:1} and \ref{ass:2},
the agents' dynamic indifference values for any bounded contingent
claim $C$ solve the following BSDE's under the minimal martingale
measure $\QQ^{\lambda,0}$ (see Proposition \ref{pro: BSDE
represantation})
\begin{eqnarray}
  -d\rho^1_t(C) &=& dX_t^{\pi^1}+\frac{1}{2}\left(\gone_t(\zeta_t^1)^2-2\zeta_t^1\phione_t\right)dt-\zeta_t^1dW_t^2, \,\,\,\,\,\,\rho^1_T(C)=-C\label{eq:BSDE1} \\
  -d\rho^2_t(C) &=& dX_t^{\pi^2}+\frac{1}{2}\left(\gtwo_t(\zeta_t^2)^2_t-2\zeta_t^2\phitwo_t\right)dt-\zeta_t^2dW_t^2, \,\,\,\,\,\,\,\rho^2_T(C)=-C\label{eq:BSDE2}
\end{eqnarray}
where $\pi^1,\pi^2\in\AA$ and $\zeta^1,\zeta^2\in\PPP_T$.

Adapting the argument lines of Section 3 of \cite{BarElk04}, we
consider a claim $C\in\linf(\FF_T)$ and introduce the BSDE
\begin{equation}\label{eq:inf-convolution dynamics}
-dF_t = f(t,\zeta_t)dt-\zeta_tdW_t^2+dX_t^{\theta},
\,\,\,\,\,\,F_T=-C
\end{equation}
where for every $t\in[0,T]$
\begin{equation}\label{eq: f_t}
    f(t,\zeta_t)=
    \frac{1}{2}\frac{\gone_t\gtwo_t}{\gone_t+\gtwo_t}\left(\zeta_t^2+(\phitwo_t-\phione_t)^2\right)-
    \frac{\zeta_t(\gone_t\phitwo_t+\gtwo_t\phione_t)+(\phitwo_t-\phione_t)^2}{\gone_t+\gtwo_t}
\end{equation}
The solution of \eqref{eq:inf-convolution dynamics} is given by a
triple $(F_t(C),\zeta_t,\theta_t)$.

\begin{lemma}\label{lem:inf-convo}
If we impose Assumptions \ref{ass:1} and \ref{ass:2} and assume
furthermore that
$\int_0^T(\phitwo_u-\phione_u)^2du\in\linf(\FF_T)$, the BSDE
\eqref{eq:inf-convolution dynamics} admits a unique solution
$(F_t(C),z_t,\theta_t)$ for every $C\in\linf(\FF_T)$, with
$(C_t)_{t\in[0,T]}$ being uniformly bounded. In addition, $F_t(C)$
is for any time $t\in[0,T]$ a convex risk measure.
\end{lemma}

\begin{proof}
First we define the process
$k_t=\frac{\gone_t\phitwo_t+\gtwo_t\phione_t}{\gone_t+\gtwo_t}$,
for $t\in[0,T]$, and note that $k\in\PPP_T^{\lambda}$. Under the
martingale measure $\QQ^{\lambda,k}$ the BSDE
\eqref{eq:inf-convolution dynamics} is written as
\begin{equation}\label{eq:inf-convolution dynamics 2}
-dF_t =
\frac{1}{2}(\Gamma_tz^2_tdt-2z_tdW^{2,k}_t)+dX_t^{\theta}+dL_t,
\,\,\,\,\,\,F_T=-C
\end{equation}
where $\Gamma_t=\frac{\gone_t\gtwo_t}{\gone_t+\gtwo_t}$,
$W^{2,k}=W_t^2+\int_0^tk_udu$ is a standard Brownian motion under
$\QQ^{\lambda,k}$, orthogonal to $W^1$ and
$L_t=\int_0^t\left(\frac{\Gamma_u}{2}(\phitwo_u-\phione_u)^2-\frac{(\phitwo_u-\phione_u)^2}{\gone_u+\gtwo_u}\right)du$.
We then observe that the BDSE
\begin{equation}\label{eq:inf-convolution dynamics 3}
-d\hat{F}_t
=\frac{1}{2}(\Gamma_tz^2_tdt-2z_td\hat{W}_t^2)+dX_t^{\theta}
\,\,\,\,\,\,\hat{F}_T=-C+L_T
\end{equation}
admits a solution $(\hat{F}_t,\hat{z}_t,\hat{\theta}_t)$, which is
in fact the dynamic risk measure $\hat{\rho}_t(C-L_T)$ induced by
the exponential performance criteria with characterization pair
$(\tone_t+\ttwo_t,k_t)$. This means that
$(\hat{F}_t-L_t,\hat{z}_t,\hat{\theta}_t)$ is a solution of
\eqref{eq:inf-convolution dynamics 2}, where
$(\hat{F}_t-L_t)_{t\in[0,T]}$ is uniformly bounded. The uniqueness
of the solution follows by Proposition \ref{pro:comparison}.

Finally, the fact that $F_t(C)$ is a dynamic convex risk measure
is a consequence of Proposition \ref{pro:comparison} and the
convexity of $f(t,\zeta_t)$ with respect to $\zeta_t$ (see also
Proposition 5.1 of \cite{Gia2006} and Proposition 5.1 of
\cite{Pen04}).
\end{proof}

The following theorem solves the optimal risk sharing problem
under forward exponential performance criteria.

\begin{theorem}\label{thm:inf-convo}
Impose Assumptions \ref{ass:1} and \ref{ass:2}, assume furthermore
that $\int_0^T(\phitwo_u-\phione_u)^2du\in\linf(\FF_T)$ and let
$(\rho_t^{1,2}(\EN),\zeta_t,\theta_t^{1,2})$ be the solution of
\eqref{eq:inf-convolution dynamics}. Then, for every $t\in[0,T]$
$$\rho_t^{1,2}(\EN)=\underset{C}{\inf}\{\rho_t^1(\EN-C)+\rho_t^2(C)\}$$
and the optimal risk sharing claims are of the form
\begin{equation}\label{eq:optimal sharing contract}
    C^*=\EN^2-\int_0^T\left(\frac{\gtwo_t}{2}\left(\frac{\gone_t\zeta_t+(\phitwo_t-\phione_t)}{\gone_t+\gtwo_t}\right)^2+
    \frac{\gone_t\zeta_t+(\phitwo_t-\phione_t)}{\gone_t+\gtwo_t}\phitwo_t\right)dt-\int_0^T\frac{\gone_t\zeta_t+(\phitwo_t-\phione_t)}{\gone_t+\gtwo_t}dW^2_t+X_T^{c,\theta}
\end{equation}
for $c\in\R$ and $\theta\in\AA^{\infty}$.
\end{theorem}
\begin{proof}
After simple calculations we get that for every processes $z$ and
$y$ and for every $t\in[0,T]$
\begin{equation}\label{eq:inequality}
f(t,z_t)\leq
\frac{1}{2}\left(\gone_t(z_t-y_t)^2-2(z_t-y_t)\phione_t\right)+
\frac{1}{2}\left(\gtwo_ty^2_t-2y_t\phitwo_t\right).
\end{equation}
Let $C\in\linf(\FF_T)$ be an arbitrarily chosen claim and
$(\rho_t^1(\EN-C),\zeta_t^1,\pi^1_t,)$ and
$(\rho_t^2(C),\zeta_t^2,\pi^2_t)$ be the solutions of the BSDE's
\eqref{eq:BSDE1} and \eqref{eq:BSDE2} with boundary conditions
$C-\EN$ and $-C$ respectively. This implies that if we set
$\tilde{\pi}_t=\pi^1_t+\pi^2_t$ and
$\tilde{\zeta}_t=\zeta_t^1+\zeta_t^2$, the triple
$(\rho_t^1(\EN-C)+\rho_t^2(C),\tilde{\zeta}_t,\tilde{\pi}_t)$ is a
solution of the following BSDE
\begin{equation}\label{eq:BSDE3}
-dC_t =
\frac{1}{2}\left(\gone_t(z_t-\zeta_t^2)^2-2(z_t-\zeta_t^2)\phione_t\right)+
\frac{1}{2}\left(\gtwo_t(\zeta^2_t)^2-2\zeta_t^2\phitwo_t\right)dt
-z_tdW_t^2+dX_t^{\theta}, \,\,\,\,C_T=-\EN
\end{equation}
for $\zeta^2$ given. Then by Proposition \ref{pro:comparison} and
inequality \eqref{eq:inequality}, we get that
$\rho_t^{1,2}(\EN)\leq\rho_t^1(\EN-C)+\rho_t^2(C)$ for every claim
$C$ and every time $t\in[0,T]$. Also for process $\zeta$
$$f_t(\zeta_t)=\frac{1}{2}\left(\gone_t(\zeta_t-\hat{z}_t)^2-2(\zeta_t-\hat{z}_t)\phione_t\right)+
\frac{1}{2}\left(\gtwo_t\hat{z}^2_t-2\hat{z}_t\phitwo_t\right)$$
and for every $t\in [0,T]$, where
$\hat{z}_t=\frac{\zeta_t\gone_t+(\phitwo_t-\phione_t)}{\gone_t+\gtwo_t}\in\PPP^{\lambda}_T$.

The next step is to observe that for the process
$$\hat{C}_t=-\int_0^t\left(\frac{\gtwo_s}{2}\hat{z}_s^2-
    \hat{z}_s\phitwo_s\right)ds+\int_0^t\hat{z}_sdW^2_t$$
the triple $(\hat{C}_t,\hat{z}_t,0)$ is the unique solution of
\eqref{eq:BSDE2} with boundary condition $\hat{C}=\hat{C}_T$,
i.e., $\hat{C}_t=\rho_t^{2}(\hat{C})$.

We also have that if $(\rho_t^1(\EN-\hat{C}),z^1_t,\pi^1_t)$ is
the solution of \eqref{eq:BSDE1} with boundary condition
$\hat{C}-\EN$, then
$(\rho_t^1(\EN-\hat{C})+\rho_t^2(\hat{C}),z_t^1+\hat{z}_t,\pi^1_t)$
is the solution of \eqref{eq:inf-convolution dynamics} with
boundary condition $-\EN$. Thanks to Lemma \ref{lem:inf-convo}, we
have that
$\rho_t^1(\EN-\hat{C})+\rho_t^2(\hat{C})=\rho_t^{1,2}(\EN)$, which
in turn means that $C^*=\EN^2-\hat{C}$ is an optimal risk sharing
contract. The fact that we can add/subtract any replicable claim
on the optimal risk sharing contract is a consequence of the
replication invariance property of the indifference valuation (see
Proposition \ref{pro:price properties}, item (iv)).
\end{proof}

\begin{remark}\label{rem:optimal contract}
We note that when $\phione_t=\phitwo_t$ for every $t\in[0,T]$,
equation \eqref{eq:inf-convolution dynamics} becomes similar to
\eqref{eq:BSDE1} and \eqref{eq:BSDE2}, where the risk aversion
process is given by $\Gamma_t$. This means that the
inf-convolution measure is a dynamic risk measures that is induced
by an agent with exponential forward performance criteria with
characteristic pair $(\Gamma_t, \phione_t)$ (see also Proposition
\ref{pro:inf-convo}).
\end{remark}

\bigskip

\appendix

\section{}

\begin{proposition}\label{pro:comparison}
Impose Assumptions \ref{ass:1} and \ref{ass:2} and assume that
$(C_t,z_t,\theta_t)$ and $(C'_t,z'_t,\theta'_t)$ are solutions of
the following BSDE's
\begin{eqnarray*}
  dC_t &=& f(t,z_t)dt+\theta_tdW_t^1+z_tdW^2_t,\,\,\,\,\,\,C_T=C \\
  dC'_t &=& f'(t,z'_t)dt+\theta'_tdW_t^1+z'_tdW^2_t,\,\,\,\,\,\,C'_T=C'
\end{eqnarray*}
where
$\underset{t\in[0,T]}{\sup}|C_t|,\underset{t\in[0,T]}{\sup}|C'_t|\in\linf(\FF_T)$,
with $C\leq C'$ a.s., $(W^1_t,W^2_t)$ is a 2-dimensional Brownian
motion, $f:\Omega\times [0,T]\times\PPP\rightarrow\R$ is given by
$$f(t,z_t)=\frac{\Gamma_t}{2}z_t^2-z_t\frac{\gone_t\phitwo_t+\gtwo_t\phione_t}{\gone_t+\gtwo_t}+
(\phitwo_t-\phione_t)^2\left(\frac{\Gamma_t}{2}-\frac{1}{\gone_t+\gtwo_t}\right)$$
and $f':\Omega\times [0,T]\times\PPP\rightarrow\R$ is a smooth
random function for which
\begin{equation}\label{eq:inequality of f}
f(t,z_t)\leq f'(t,z_t),\,\,\,\text{a.s.}
\end{equation} Then, $C_t\leq C'_t$,
a.s.~for every $t\in[0,T]$.
\end{proposition}

\begin{proof}
\begin{eqnarray*}
  C_t-C'_t -(C_0-C'_0) &=& \int_0^t(f(s,z'_s)-f'(s,z'_s))ds + \int_0^t(z_t-z'_t)dW^2_s \\
   & + & \int_0^t(f(s,z_s)-f(s,z'_s))ds +\int_0^t(\theta_t-\theta'_t)dW^1_s
\end{eqnarray*}
We then observe that
\begin{eqnarray*}
  f(t,z_t)-f(t,z'_t) &=& \frac{\Gamma_t}{2}(z_t^2-(z_t')^2)-(z_t-z'_t)\frac{\gone_t\phitwo_t+\gtwo_t\phione_t}{\gone_t+\gtwo_t} \\
    &=& (z_t-z'_t)K_t
\end{eqnarray*}
where
$K_t=\frac{\Gamma_t}{2}(z_t+z'_t)-\frac{\gone_t\phitwo_t+\gtwo_t\phione_t}{\gone_t+\gtwo_t}$.
Now $K\in\PPP^{\lambda}_T$ by Lemma \ref{lem:bounds} and the
uniform boundness of $\gone$ and $\gtwo$. Hence, $\QQ^{\lambda,K}$
is a martingale measure and therefore  $ C_t-C'_t
-(C_0-C'_0)-\int_0^t(f(s,z'_s)-f'(s,z'_s))ds$ is a true
$\QQ^{\lambda,K}$-martingale (see also the proof of Theorem
\ref{thm:robust repre}). Taking expectation under
$\QQ^{\lambda,K}$ and exploiting the assumed inequality
\eqref{eq:inequality of f} gives the intended inequality of the
solutions.
\end{proof}

\bigbreak

\end{document}